\documentclass[
  aip,
  jcp,
  amsmath,amssymb,
  reprint,
  floatfix
]{revtex4-1}
\usepackage{graphicx}% Include figure files
\usepackage{dcolumn}% Align table columns on decimal point
\usepackage{bm}% bold math
\usepackage{hyperref}
\usepackage{siunitx}
\DeclareSIUnit{\atmosphere}{atm}
\DeclareSIUnit\angstrom{\text{Å}}
\DeclareSIUnit\bar{bar}
\usepackage{mathptmx}
\usepackage{etoolbox}
\makeatletter
\def\@email#1#2{%
 \endgroup
 \patchcmd{\titleblock@produce}
  {\frontmatter@RRAPformat}
  {\frontmatter@RRAPformat{\produce@RRAP{*#1\href{mailto:#2}{#2}}}\frontmatter@RRAPformat}
  {}{}
}%
\makeatother

\begin{document}
%\preprint{AIP/123-QED}

\title{Robust Kirkwood-Buff Inversion in Complex Mixtures via Reciprocal-Space Methods}

\author{R.Busselez}
 \email{remi.busselez@univ-lemans.fr}
 \affiliation{Institut des Molécules et Matériaux du Mans, UMR CNRS 6283,
Le Mans université, 72085 Le Mans, France} 

\date{\today}% It is always \today, today,

\begin{abstract}
Understanding the relationship between microscopic structure and macroscopic thermodynamic properties is a central challenge in the study of complex fluids. The Kirkwood-Buff (KB) theory offers an elegant and powerful framework for bridging this gap by relating integrals over pair correlation functions to measurable thermodynamic quantities. In multicomponent systems, KB integrals connect directly to derivatives of thermodynamic potentials, including chemical potentials derivatives, partial molar volumes, and isothermal compressibility. While several computational methods exist to estimate KB integrals from molecular simulations, their application often demands careful treatment of finite-size effects and explicit extrapolation to the thermodynamic limit. Recently, alternative strategies based on the analysis of partial structure factors in reciprocal space have been proposed. Unlike real-space approaches, reciprocal-space methods avoid the additional truncation artifacts associated with direct integration or fluctuations in subensemble. They evaluate density fluctuations across the entire simulation box, fully accounting for periodic boundary conditions rather than relying on subdomains. As a result, these methods offer a compelling alternative, providing enhanced numerical stability for estimating KB integrals in complex mixtures. In this work, we extend, compare and validate these methods using binary and quaternary Lennard-Jones mixtures, as well as realistic molecular systems such as hexane-ethanol, water-urea and aqueous NaCl mixtures. Our results provide practical guidelines for computing KB integrals and associated thermodynamic properties from canonical ensemble simulations, including recommendations on reciprocal-space extrapolation, uncertainty estimation and linear algebra formulations of thermodynamic derivatives.
\end{abstract}
\maketitle
\section{Introduction}
Establishing a rigorous connection between molecular-level structure and macroscopic thermodynamic behavior remains a central challenge in the study of complex fluids and mixtures \cite{debenedetti1987tjocpa, debenedetti1987tjocp,debenedetti1986cpl,kjelstrup2014ansnn, schnell2011jpcb}. The Kirkwood–Buff (KB) theory\cite{kirkwood1951tjocp} provides a powerful and elegant framework for bridging this gap by linking integrals over microscopic pair correlation functions to thermodynamic derivatives\cite{ben-naim2006,newman1994csr}, such as partial volumes, isothermal compressibility, and chemical potential gradients. While the KB theory has been extensively applied in both experimental\cite{alimansoori1980fpe, alimansoori1980fpe, ben-naim1990pac, ben-naim2008tjocp,chitraMolecularAssociationSolution2002, almasyStructureAqueousSolutions2008, almasySmallangleNeutronScattering2004, pereraEvaluationKirkwoodBuffIntegrals2005, pereraKirkwoodBuffIntegralsAqueous2006, shulginKirkwoodBuffIntegralsAqueous1999,seo2024jpcb} and computational contexts \cite{venetsanos2022m,fingerhut2019fpe,galata2018fpe, lbadaoui-darvas2019jpcb, mukherjiPreferentialSolvationTriglycine2012, pereraComparativeMolecularDynamics2011, perera2022jcpa, petris2019jpcb}, the accurate evaluation of Kirkwood–Buff integrals (KBIs) from molecular simulations presents persistent difficulties. These arise primarily from finite-size effects \cite{braten2021jcim, dawass2018ms, heidari2018mp, romanFluctuationsSmallHarddisk1999, simon2022jcp, stromSizeShapeEffects2017, villamainaThinkingOutsideBox2014} and the inherent differences between closed simulation ensembles and the open, infinite systems assumed in KB theory. Consequently, a variety of techniques have been developed to overcome these challenges, including real-space integration methods with finite-size corrections\cite{kruger2013jpcl,gangulyConvergenceSamplingKirkwood2013,dawass2018ms,dawass2019fpe}, spatial block analysis\cite{sevilla2022jcp,heidari2018mp,cortes-huerto2016tjocp,heidari2018e}, and more recently, reciprocal space approaches based on partial structure factors\cite{nichols2009pre,nichols2015iecr,cheng2022jcp,rogersExtensionKirkwoodBuffTheory2018,schmidComputingChemicalPotentials2023}.
In this work, we focus on advancing and validating reciprocal-space methodologies for the extraction of KB integrals and associated thermodynamic properties. Building on the theoretical framework proposed by Nichols \textit{et al.}~\cite{nichols2009pre}, we employ and extend the use of Fourier components of direct correlation functions, whose smoother low-$q$ behavior enables more reliable extrapolation to the thermodynamic limit. This approach significantly reduces the system size required for accurate calculations, offering practical advantages over traditional real-space methods. To evaluate the performance and generality of the proposed methodology, we apply it to a diverse set of systems with increasing complexity: from binary Lennard-Jones mixtures to realistic molecular liquids such as hexane–ethanol, aqueous urea, and sodium chloride solutions. These systems present distinct challenges, including strong composition fluctuations, molecular complexity, and long-range electrostatic interactions. Our analysis includes the determination of KB integrals, activity coefficients derivatives, thermodynamic and excess mixing properties, and where applicable, comparison with both experimental data and established models.
By systematically analyzing these systems, we aim to demonstrate the robustness and accuracy of reciprocal-space KB analysis and to provide practical guidelines for its application to a wide range of multicomponent fluid systems.

\section{Theory}
\subsection{Kirkwood Buff Solution Theory}
In this section we will summarize general results obtained in the seminal work of \citealt{kirkwood1951tjocp}, \citealt{oconnell1971mpa} and \citealt{nichols2009pre}. As our main goal is to connect atomic simulations to fluctuation solution theory using Kirkwood Buff approach, the development will be given on a molecular or atomic number basis rather than molar basis (molarity or molality).
We first consider an open n-multicomponent system in a volume \(V\) at temperature \(T\) in contact with an infinite reservoir of particles, each species $i$ at equilibrium with a fixed chemical potential $\mu_i$.
The composition fluctuation can be directly related through the fluctuation theorem to the second derivative of the grand potential $\Omega$ in respect to chemical potentials or to the first derivative of the average number of species with respect to the chemical potential \cite{newman1994csr,kirkwood1951tjocp}:
\begin{eqnarray}
  \left(-\frac{\partial^2 \Omega}{\partial \mu_{i} \partial \mu_{j}}\right)_{V,T,N_{\gamma}}&=&\left(\frac{\partial \langle N_{i}\rangle}{\partial \mu_{j}}\right)_{V,T,N_{\gamma}}=\left(\frac{\partial \langle N_{j}\rangle}{\partial \mu_{i}}\right)_{V,T,N_{\gamma}} \nonumber \\ 
  &=& \frac{1}{kT}\left(\langle N_{i}N_{j}\rangle - \langle N_{i}\rangle\langle N_{j}\rangle\right) \nonumber\\
  &=&\frac{1}{kT} \langle\delta N_{i}\delta N_{j}\rangle
  \label{eq:fluct_gc}
\end{eqnarray}
Where $N_{\gamma}$ indicates a constant number of particles other than $i$ or $j$ during differentiation and bracket the ensemble averaging.
Introducing the instantaneous fluctuation of species i number \(\delta N_{i}\) as :
\begin{equation}
\delta N_{i}=N_{i}-\langle N_{i}\rangle=\int_V \left(\rho_{i}(\bm{r})-\langle\rho_{i}\rangle\right) d\bm{r}= \int_V \delta \rho_{i} (\bm{r}) d\bm{r}
\end{equation}
The composition fluctuation between $i$ and $j$ species can be rewritten via the double volumic integral of the density fluctuation correlation function
\begin{equation}
  \left\langle \delta N_i \delta N_j \right\rangle = \int_V \int_V \Bigl\langle \delta\rho_i(\bm{r_1}) \delta\rho_j(\bm{r_2})\Bigr\rangle d\bm{r_1} d\bm{r_2}
    \label{eq:densfluccorr}
\end{equation}

Introducing $h^{(2)}_{ij}(\bm{r_1},\bm{r_2})$ the partial pair correlation function between species $i$ and $j$ and $g^{(2)}(\bm{r_1},\bm{r_2})$ the corresponding partial pair distribution function both related via\cite{hansenTheorySimpleLiquids2013} $h^{(2)}_{ij}(\bm{r_1},\bm{r_2})= g^{(2)}_{i,j}(\bm{r_1},\bm{r_2})-1$

The fluctuation composition between species $i$ and $j$ can be rewritten in terms of local density of species and partial pair correlations functions

%\begin{equation}
\begin{multline}
  \left\langle \delta N_i \delta N_j \right\rangle =\int_V\int_V \Bigl(\rho_i(\bm{r_1})\rho_j(\bm{r_2}) h^{(2)}_{ij}(\bm{r_1}, \bm{r_2}) \\
  + \rho_{i}(\bm{r_1}) \delta_{i,j} \delta(\bm{r_1} -\bm{r_2})\Bigr) d\bm{r_1} d\bm{r_2}
\end{multline}
%\end{equation}
With $\delta_{i,j}$ the standard Kronecker symbol, which equals 1 when $i = j$ and 0 otherwise and $\delta(\bm{r})$ the three-dimensional Dirac delta distribution.
When species are represented as single point particles, the local microscopic density of a given species and its local fluctuation around an average value can be described using the Fourier components of the microscopic density \(\rho_i({\bm{q}})\) \cite{hansenTheorySimpleLiquids2013,nichols2009pre}:

\begin{equation}
  \begin{aligned} 
  \rho_i(\bm{r}) &= \frac{1}{V} \sum_{\bm{q}} \rho_i(\bm{q}) \exp(\imath\bm{q}\bm{r})\\
  \delta \rho_i(\bm{r}) &= \frac{1}{V} \sum_{\bm{q}\neq\bm{0}} \rho_i(\bm{q}) \exp(\imath\bm{q}\bm{r})\\
  \rho_{i}(\bm{q}) &= \sum_{i} \exp(-\imath \bm{q}\bm{r_{i}})
  \end{aligned}
  \label{eq:fourier_coeff_com}
\end{equation}
Introducing the partial structure factor between species $i$ and $j$ in the same manner as \cite{hansenTheorySimpleLiquids2013}
\begin{equation}
  S_{ij}(\bm{q})=\left\langle \frac{1}{N} \rho_i(\bm{q})\rho_j(\bm{-q})
\right\rangle
\label{eq:sqadim}
\end{equation}
With $N=\sum_{i} N_{i}$ the total number of particles in the open system and $\langle N\rangle$ the mean value.
Combining Eq.~\eqref{eq:densfluccorr} and \eqref{eq:sqadim}, we retrieve the well-known relationship between particle number fluctuations and partial structure factor in grand-canonical ensemble at thermodynamic limit\cite{hansenTheorySimpleLiquids2013,nichols2009pre}
\begin{equation}
  \lim_{\bm{q}\to\bm{0}} S_{ij}(\bm{q}) =\frac{1}{\left\langle N\right\rangle} \left\langle\delta N_i\delta N_j \right\rangle = \frac{kT}{\left\langle N\right\rangle} \left(\frac{\partial \left\langle N_i\right\rangle}{\partial \mu_j}\right)_{T,V,N_{\gamma}}
  \label{eq:s0}
\end{equation}

In grand canonical open ensemble, the determination of partial structure factors between all the different species constituting the system permits to define \(\bm{S(q)}\) a square symmetrical matrix of rank $n$ formed by the partial structure factor values for a given $\bm{q}$ wavevector, this matrix is related to the matrix \(\bm{H(q)}\) composed by the Fourier components of the pair correlation function between different species at wavevector $\bm{q}$ \cite{hansenTheorySimpleLiquids2013, oconnell1971mpa,nichols2009pre} such that 
\begin{equation}
  \bm{S(q)} = \bm{X}  + \rho \bm{X} \bm{H(q)}\bm{X}
  \label{eq:indirectcf}  
\end{equation}
Where \(\rho=\frac{\langle N\rangle}{V}\) the total number density and \(\bm{X}\) is a diagonal matrix containing the number fraction of distinct species \(\bm{X}_{i,j}=\delta_{i,j} \frac{N_i}{N}\).

%The symmetric $\bm{H(q)}$ matrix is constituted by all the fourier component at wavevector $\bm{q}$ of the different partial pair correlation function between the n species.

Towards low q limit, fourier components of pair correlation functions converge to Kirkwood-Buff integral elements\cite{oconnell1971mpa,kirkwood1951tjocp,nichols2009pre,hansenTheorySimpleLiquids2013} defined as
\begin{equation}
  \lim_{q\to 0} \bm{H}_{ij}(\bm{q}) = \bm{G}_{i,j} = V\left[\frac{\langle N_{i} N_{j}\rangle - \langle N_{i}\rangle \langle N_{j}\rangle}{\langle N_{i}\rangle \langle N_{j}\rangle} -\frac{\delta_{i,j}}{\langle N_{i}\rangle}\right]
  \label{eq:kbielements}
\end{equation}
With $\delta_{i,j}$ the standard Kronecker symbol, which equals 1 when $i = j$ and 0 otherwise.
Those elements are also related in grand-canonical ensemble to the double volumic integrals of pair correlation functions or to the simple integral of the radial distribution function $g_{ij}(r)$ through the assumption of an isotropic system at thermodynamic limit permitting the variable change $\bm{r}=\bm{r_2}-\bm{r_1}$, leading to the following expressions of KB integrals via pair correlation function integral or radial distribution function:
\begin{eqnarray}
  \bm{G}_{i,j}&=&\frac{1}{V}\int_V\int_V \left(g_{i,j}^{(2)}(\bm{r_1},\bm{r_2})-1\right) d{\bm{r_1}} d\bm{r_2} \label{eq:kbi_double_sum} \\
  &=&\int (g_{i,j}(r)-1) 4\pi r^2 dr \label{eq:kbi_rdf_sum} 
\end{eqnarray}

Following the generalization of Ornstein-Zernike theory for multicomponent system\cite{hansenTheorySimpleLiquids2013}, the matrix $\bm{H(q)}$ containing the Fourier components of pair correlation at a given wavevector can be related to the matrix of Fourier components of direct pair correlation at the same wavevector \(\bm{C(q)}\) \cite{hansenTheorySimpleLiquids2013,oconnell1971mpa,nichols2009pre} and then are directly connected to the matrix $\bm{S(q)}$ via a matrix inversion\cite{nichols2009pre,oconnell1971mpa,hansenTheorySimpleLiquids2013}.
\begin{equation}
  \rho \bm{C(q)}= \bm{X^{-1}} - \bm{S(q)^{-1}}
  \label{eq:directcf}
\end{equation}

Focussing on the $\bm{q}\to 0^+$ limit of $\bm{S(q)}$ we obtain the symmetrical square \(\bm{S}\) matrix whose elements are given by the thermodynamic derivatives \eqref{eq:s0} and related to the low-q limit of the fourier transform values of direct pair correlations \eqref{eq:directcf} via a matrix inversion\cite{nichols2009pre,oconnell1975jsc,oconnell1995fpe}.

\citealt{kirkwood1951tjocp} demonstrate that the elements of  symmetrical \(\bm{S}\) matrix containing the partial derivative of particle number of one species with respect to the chemical potential of another species in grand-canonical ensemble (constant volume, temperature, chemicals potentials and constant number of particles not involved in the derivative) are related to the derivatives of the chemicals potentials with respect to number of particles in canonical ensemble (constant volume, temperature and number of particles) using Legendre transform.
It is then possible to express the elements of \(\bm{A}\) matrix through the Moore-Penrose pseudo-inverse of \(\bm{S}\) matrix \cite{kirkwood1951tjocp,nichols2009pre,oconnell1971mpa,oconnell1975jsc,ben-naim2006}
\begin{equation}
  \bm{A}_{i,j}=\left(\bm{S}^{-1}\right)_{i,j}=\frac{N}{k_B T}\left(\frac{\partial \mu_{i}}{\partial N_{j}}\right)_{T,V,N_{\gamma}}
\end{equation}

The above expression of thermodynamic derivative is obtained in the canonical ensemble, in order to obtain analogous expressions in the Gibbs ensemble, another Legendre transform is employed on canonical relations, using Gibbs-Duhem relationship and thermodynamic identities conducts to the following expression\cite{kirkwood1951tjocp,ben-naim2006,oconnell1971mpa}
\begin{eqnarray}
  \bm{D}_{i,j}&=&\frac{N}{k_B T}\left(\frac{\partial \mu_{i}}{\partial N_{j}}\right)_{T,P,N_{\gamma}} \nonumber\\
  &=&\frac{N}{k_B T}\left(\frac{\partial \mu_{i}}{\partial N_{j}}\right)_{T,V,N_{\gamma}} -\rho V_{i}\rho V_{j}\frac{\kappa_{ig}}{\kappa_T} \label{eq:npt_derivative}
\end{eqnarray}
With \(V_{i}=\left(\frac{\partial V}{\partial N_{i}}\right)_{N_{\gamma\neq i}, P,T}\) the partial volume of species \(i\), \(\kappa_T=-\frac{1}{V}\left(\frac{\partial V}{\partial P}\right)_{T,N}\) the isothermal compressibility of the system and \(\kappa_{ig}=\frac{1}{\rho k_B T}\) the thermal compressibility of a perfect gas at same density and temperature.
Following the work of refs \onlinecite{ben-naim2006,oconnell1971mpa,nichols2009pre}, the elements of the $D$ matrix \eqref{eq:npt_derivative} can be expressed using linear algebra formalism relating \(\bm{A}\) matrix, \(\bm{v}\) a column vector of partial volumes, \(\bm{x}\) a column vector containing the average number fraction of molecules and \(\rho\) the total number density
\begin{equation}
  \bm{v} = \frac{1}{\rho}\frac{\bm{A} \bm{x}}{\bm{x}^{T} \bm{A} \bm{x}}
\end{equation}
The isothermal compressibility is expressed by
\begin{equation}
  \kappa_T=\kappa_{ig} \left(\bm{x}^T \bm{A} \bm{x}\right)^{-1}
\end{equation}
Leading to the expression of the following \(\bm{D}\) matrix whose elements are the derivative of chemical potentials with respect to particles number in the Gibbs ensemble
\begin{equation}
  \bm{D} = \bm{A} -\rho^2 \left(\bm{x}^T \bm{A}\bm{x}\right) \left(\bm{v} \bm{v}^T\right)=\bm{A} - \frac{(\bm{A}\bm{x})(\bm{x}^T\bm{A})}{\bm{x}^T \bm{A} \bm{x}}
\end{equation}

Chemicals potentials of species in a multi-component system are expressed via the chemical potential in Gibbs ensemble for pure components and the activity coefficient of species in the following manner
\begin{equation}
  \mu_i(P,T,x_i) = \mu_i^0(P,T) + k_BT \ln(x_i \gamma_i(P,T, x_i))
  \label{eq:chempotmole}
\end{equation}
where \( \mu_i^0(P,T) \) is the chemical potential of the pure component at fixed $T$ and $P$, \( \gamma_i(P,T,x_i) \) is the activity coefficient of species \( i \) at fixed \( T \), \( P \) and species $i$ molar fraction \(x_i\).

Relating the elements of the matrix \(\bm{D}\) to the expression \ref{eq:chempotmole} implies a transformation from a particle (or molar) number basis to a particle (or molar) fraction basis.
The ubiquitous constraint on the fraction sum \(\sum_{i}x_{i}=1\), involve that a change in a fraction of species \(j\) is related to a change in the fractions of the other \(n-1\) species.
To resolve this issue, it is common to select a specific species \(k\) as a reference species, which allows the \(\Sigma\) constraint to be maintained during differentiation.
Therefore, the derivative of a thermodynamic parameter with respect to the molar fraction of species \(j\) can be expressed as a change in the parameter with respect to the number of species \(j\) considered, along with the adjustment of the parameter due to a variation in the number of species \(k\), ensuring that the constraint \(\Sigma\) remains intact. The thermodynamic factor \(\Gamma_{i,j}\) is related to the derivative of the chemical potential of one species with respect to fraction of another species \cite{ben-naim2006,jonah1994fpe,taylor1991cec}.
\begin{equation}
  \bm{\Gamma}_{i,j}=\frac{x_i}{k_B T}\left(\frac{\partial\mu_i}{\partial x_j}\right)_{T,P,\Sigma}=\delta_{i,j}+\left. x_i\frac{\partial\ln(\gamma_i)}{\partial x_j}\right\vert_{T,P,\Sigma}
\end{equation}
With $\delta_{i,j}$ the standard Kronecker symbol, which equals 1 when $i = j$ and 0 otherwise.
The choice of a reference species $k$ leads to the following relationship between chemical potential derivative on a number basis and chemical potential on a fraction basis\cite{ben-naim2006,jonah1994fpe,taylor1991cec,fingerhut2020mp}
\begin{equation}
  \bm{\Gamma}_{i,j}= x_i \frac{N}{k_BT} \left[\left(\frac{\partial \mu_i}{\partial N_j}\right)_{T,P,N_{\neq j}} -\left(\frac{\partial \mu_i}{\partial N_k}\right)_{T,P,N_{\neq k}} \right]
  \label{eq:num2mol}
\end{equation}
%The introduction of the \(\Sigma\) constraint, and the choice of reference species imply that \(\Gamma\) matrix is of rank \((n-1)\).
Introducing the matrix \(\bm{I_0}\) as the addition of an identity matrix of rank \(n\) and a \(n\) null matrix whose \(k^{th}\) row value is \(-1\) such that \(\bm{I_0}=\bm{I_n} + \bm{Z_k}\). The non symmetric \(\bm{\Gamma}\) matrix of rank \(n\) is related to the symmetric \(\bm{D}\) matrix of rank $n$ through
\begin{equation}
  \bm{\Gamma} = \bm{X} \bm{D}\bm{I_0}
  \label{eq:gamma_matrix}
\end{equation}
With the \(k^{th}\) column of \(\bm{\Gamma}\) equal to zero, the Gibbs-Duhem relationship imply that \(\sum_i \Gamma_{ij}=0\), combining both observations the following relation between coefficients are obtained \(\bm{\Gamma}_{i,j} / x_i - \bm{\Gamma}_{j,i} / x_j + (\bm{\Gamma}_{k,i} - \bm{\Gamma}_{k,j}) / x_k = 0\).
Thus the \(k^{th}\) row of the matrix can be expressed by a linear relation between other rows, it is then possible to define from the non-symmetric matrix  \(\bm{\Gamma}\) of rank \(n\) a non-symmetric square matrix \(\bm{\tilde{\Gamma}}\) of rank \((n-1)\) by removing the \(k^{th}\) row and column.

The elements of \(\bm{\tilde{\Gamma}}\) matrix are also called Darken factors\cite{gazzilloStabilityFluidsMore1994, guevara-carrion2021sr, krishnaDarkenRelationMulticomponent2005, liuPredictiveDarkenEquation2011, vignes1966iecf} which represents gradients of chemical potentials with respect to molar fraction, they are hence directly related to Onsager coefficients coupling fluxes through gradients of chemical potential with respect to components fraction, among them the Maxwell-Stefan diffusion coefficients\cite{guevara-carrion2021sr, nichols2015iecr,van-brunt2023ea,schoen1984mp,simon2022jcp,krishnaDarkenRelationMulticomponent2005,liuPredictiveDarkenEquation2011,vignes1966iecf} or Soret coefficients\cite{gittus2023pccp,a.rudaniAnalyzingConcentrationdependentSoret2025}.

\subsection{Evaluation of Kirkwood Buff Integral}

All the thermodynamic derivatives expressed in the above section can be extracted from the Kirkwood Buff integrals (KBIs) using the inversion of KB relationships. Theoretically KBIs can be directly calculated using the radial distribution function as described by Eq.\eqref{eq:kbi_rdf_sum} or from compositions fluctuations in grand-canonical ensemble. However in practice several challenges arise when evaluating KBIs, a first complication is the practical limitations for sampling open systems. In molecular dynamics grand-canonical ensemble is difficult to reach and only monte-carlo simulations are able to sample accurately the grand-canonical ensemble. The KBIs are typically computed from closed finite ensembles (NVT or NPT) subject to periodic boundary conditions, whereas the KB theory is formulated in open ensemble and infinite systems. This mismatch introduces a first series of finite-size effects\cite{matubayasiThermodynamicsHydrationShell1996,salacuseFinitesizeEffectsMolecular1996,salacuseFinitesizeEffectsMolecular1996a,rogersExtensionKirkwoodBuffTheory2018}, affecting the pair correlation functions and consequently the RDFs\cite{lebowitzLongRangeCorrelationsClosed1961}. As a result, local density fluctuations in subvolume or RDFs obtained from closed pseudo-infinite ensembles must be corrected \cite{ben-naim2006,simon2022jcp,gangulyConvergenceSamplingKirkwood2013,milzettiConvergenceKirkwoodBuff2018,verlet1968pra,siepmannFinitesizeCorrectionsChemical1992,romanFluctuationsEquilibriumHarddisk1997,romanFluctuationsSmallHarddisk1999}.
Additionally, others finite-size effects arise when performing a finite volumic integral at a finite cut-off radius rather than integrating to the thermodynamic limit\cite{salacuseFinitesizeEffectsMolecular1996a, salacuseFinitesizeEffectsMolecular1996,rogersExtensionKirkwoodBuffTheory2018}.
To address these issues, two main techniques using corrections have been developed. The first is based on spatial block analysis method (SBAM) where the simulation box is subdivised into smaller subvolumes particle fluctuations are calculated within those open and connected subvolumes, although the overall simulation box is finite\cite{heidari2018mp,cortes-huerto2016tjocp,heidari2018e,rovereGasliquidTransitionTwodimensional1990}, using thermodynamics of small systems \cite{hill2013, schnell2011jpcb,bedeaux} it remains possible to extrapolate the obtained results towards an infinite-volume limit taking into account the subvolume size effect on fluctuations\cite{rovereGasliquidTransitionTwodimensional1990,cortes-huerto2016tjocp, sevilla2022jcp}.
A second approach involves the reformulation of the infinite double volume integral in Eq.\eqref{eq:kbi_double_sum} to a finite domain, explicitly accounting for the excluded volume effect that arises when integrating over a limited region. Thus reformulating the relationship between KB integrals and RDF as:
\begin{equation}
  G_{ij}(R) = \int_0^{2R} w(r,R) (g_{ij}(r) -1) 4\pi r^2 dr \label{eq:KBI_R}
\end{equation}
where the weighting function $w(r,R)$ depends on the subvolumes geometry\cite{bedeaux, dawass2018ms, kruger2013jpcl, simon2022jcp}.
By performing the integration for different subdomain sizes $R$, the resulting $G_{ij}(R)$ curve can be extrapolated to $R\to\infty$.
A further challenge in this integration procedure is the amplification of the long-range part of the RDF by the $r^2$ weighting factor. This can lead to poor convergence, particularly when long-range correlations are naturally present, or when RDF are not corrected from closed system finite size effects\cite{gangulyConvergenceSamplingKirkwood2013,kruger2013jpcl}. The convergence and accuracy of the KBIs obtained via RDF integrals remains notoriously difficult even when corrections are applied and advanced sampling techniques are used to improve the statistics at large distances\cite{borgis2013mp, coles2019jcp, rotenberg2020jcp}.

As an alternative to real-space determination of KBIs, reciprocal-space methods connect partial structure factors to KBIs via Eq.~\eqref{eq:indirectcf}. In this approach, the structure factor defined in Eq.~\eqref{eq:sqadim} is computed from the discrete Fourier series of atomic positions, evaluated at reciprocal lattice vectors compatible with the geometry of the simulation box, as described in Eq.~\eqref{eq:fourier_coeff_com}.

When applied to finite-size systems with periodic boundary conditions, there are several advantages to computing structure factors using the discrete Fourier transform of the density fluctuation spectrum~\cite{horbach1996pre,villamainaThinkingOutsideBox2014,rogersExtensionKirkwoodBuffTheory2018,nichols2009pre}, rather than through the more commonly used Fourier transform of the radial distribution function (RDF), as given by Eq.~\eqref{eq:sqcontinuous}~\cite{salacuseFinitesizeEffectsMolecular1996,salacuseFinitesizeEffectsMolecular1996a}:

\begin{equation}
S_{ij}(q) = x_i \delta_{i,j} + x_i x_j\rho \int_0^{\infty} 4\pi r^2 (g_{ij}(r)-1) \frac{\sin(qr)}{qr} \, dr
\label{eq:sqcontinuous}
\end{equation}
Specifically, using Eqs.~\eqref{eq:fourier_coeff_com} and~\eqref{eq:sqadim}, the Fourier components of density fluctuations are sampled at discrete lattice vectors spanning the full volume of the simulation cell. By construction, this formalism respects periodic boundary conditions and avoids the artifacts associated to finite spherical integration in finite real space. As a result, this method substantially mitigates finite-size effects associated with volumic integral~\cite{horbach1996pre,nichols2009pre,sevilla2022jcp,villamainaThinkingOutsideBox2014}.
Moreover, computing $S_{ij}(q)$ from Fourier components of density fluctuations ensure that the result is naturally invariant under uniform shifts in $g_{ij}(r)$ or in the asymptotic value of local density $lim_{r\to\infty} \rho(r)$ relative to an arbitrary origin\cite{rogersExtensionKirkwoodBuffTheory2018,nichols2009pre}.
However, as noted by Rogers\cite{rogersExtensionKirkwoodBuffTheory2018}, the structure factors are calculated in a finite system using a closed ensemble, so the difficulty in reaching the thermodynamic limit stems from ensemble-size dependence. In this sense, finite-size effects do not vanish entirely but persist due to ensemble constraints, particularly when the system's correlation length approaches the box size~\cite{cheng2022jcp,rogersExtensionKirkwoodBuffTheory2018}.

Nevertheless, the $q\to 0^+$ limit required for the determination of $\bm{S}$ matrix and calculation of KBIs, cannot be accessed directly in simulation. Instead, an analytic function must be used to describe the low-q region of the structure factor, and the thermodynamic limit value at $q\to 0^+$ must be extrapolated\cite{cheng2022jcp,nichols2009pre,nichols2015iecr,rogersExtensionKirkwoodBuffTheory2018}.
The so-called $S_0$ method\cite{cheng2022jcp} enables the estimation of KBIs and related thermodynamic derivatives from simulation data using the partial structure factor coupled to an Ornstein-Zernike approximation\cite{ornstein} $S_{ij}(q)=\frac{S_{ij}(0)}{1+\xi_{ij}^2 q^2}$  used to fit the low-q region and extrapolate the $q\to 0^+$ value.

In our view, this method presents two closely related pitfalls. First, the structure factor is directly linked to the Fourier transform of the pair correlation function, which decays slowly to zero at long distances in real space. As a result, the values of $S_{ij}(q)$ can exhibit substantial fluctuations in the low-$q$ region of the spectrum, rendering the extrapolation to $q = 0$ potentially unreliable and highly sensitive to noise. This sensitivity can introduce significant uncertainty in the estimated macroscopic values. 

Second, the commonly employed Ornstein-Zernike approximation for extrapolating $1/S_{ij}(q)$ is typically truncated at order $q^2$. While this quadratic truncation is often appropriate for systems with short-range interactions and within a narrow $q$-range near $q \to 0$, it becomes inadequate for systems with longer-range interactions or for structure factor data extending to higher $q$-values. In such cases, the asymptotic expansion requires the inclusion of higher-order $q$-terms.

As pointed out by \citealt{oconnell1971mpa}, the direct pair correlation function in real space is of shorter range than the (indirect or total) pair correlation function. Consequently, the Fourier transform of the direct correlation function converges to its asymptotic low-q value at higher wavevectors values than that of the fourier transform of pair correlation function. This property allows for a more monotonic description of the low-wavevector limit, improving the reliability of asymptotic expansion leading to a more robust macroscopic extrapolations. This insight forms a central aspect of the work by \citealt{nichols2009pre}, who explored the analysis analytic expansion for the Fourier components of the direct correlation function, using the Ornstein-Zernike equation truncated up to the $q^2$ term.

It is also important to recall that the long-range part of the direct correlation function in real space is linked to the asymptotic decay of the interaction potential\cite{hansenTheorySimpleLiquids2013,nichols2009pre,wedberg2010ms}, such that $c_{ij}(r)\propto -\frac{\phi_{ij}(r)}{kT}$, where $\phi_{ij}(r)$ is the interaction potential between particles $i$ and $j$. The standard low-wavevector expansion of $C_{ij}(q)$ in even powers of $q$ is valid for systems with short-range potentials, those that vanish beyond a cutoff or decay exponentially \cite{hansenTheorySimpleLiquids2013,enderby1965pps,evansAsymptoticDecayCorrelations1994a,wedberg2010ms}. However, for potentials of longer range with slower decay in real space, such as the $r^{-6}$ dispersion term, a $\vert\bm{q}\vert^3$ contribution appears in the low-q expansion of $C_{ij}(q)$\cite{enderby1965pps, carvalho1994jpcm, woodhead-galloway1968jpcssp,hansenTheorySimpleLiquids2013,rogersExtensionKirkwoodBuffTheory2018}. As an example, in a Lennard-Jones monoatomic fluid, the asymptotic low-q expansion of $C(q)$ can be described by an Ornstein-Zernike expansion in even powers of $q$ supplemented by a $\vert \bm{q}\vert^3$ term that account for the dispersive component of the van der Waals interaction \cite{rovere1979pmb, kjellander2005jpamg, salmon2006jpcma}.

\section{Methodology}
\subsection{System preparation and obtention of trajectories}

We apply the following methodology across all systems studied. First, we construct the molecular structures and assign force fields using Parmed \cite{shirtsLessonsLearnedComparing2017} and Foyer \cite{kleinFormalizingAtomtypingDissemination2019}. The specific force fields used for each system are detailed in the corresponding results sections. Initial configurations are generated with Packmol \cite{martinezPACKMOLPackageBuilding2009}, after which the force fields are applied using Parmed. Molecular dynamics (MD) simulations are carried out using OpenMM v7.7 \cite{eastmanOpenMM7Rapid2017}, following the procedure outlined below. We begin by equilibrating the system in the NPT ensemble, maintaining the target pressure with a stochastic barostat and regulating temperature using a Langevin thermostat with the VRORV (BAOAB) integration scheme \cite{leimkuhler2013tjocp}. Interatomic potentials are truncated at \SI{1.2}{\nano\meter}, with a switching function applied between \qty{1.1}{\nano\m} and \qty{1.2}{\nano\m}, and energy corrections are applied accordingly. All bonds in molecular compounds are constrained. Coulomb interactions (when presents) are computed using the Particle Mesh Ewald (PME) method \cite{darden1993jcpa}, with a \qty{1.2}{\nano\m} cut-off and a force tolerance of $10^{-4}$. Simulations use a \qty{2}{\femto\s} timestep and run for \qty{20}{\nano\s}, with thermodynamic data recorded every 250 timesteps during the final \qty{10}{\nano\s}.

Subsequently, a simulation is conducted in the NVT ensemble, starting from the final NPT configuration. The simulation cell is resized to the average box dimensions from the last \SI{10}{\nano\s} of NPT run, and atoms are remapped accordingly. This NVT run spans \qty{25}{\nano\s}, with the first \qty{5}{\nano\s} discarded. Configurations are saved every \qty{2}{\pico\s}, yielding 10000 configurations over the final \qty{20}{\nano\s}.

\subsection{Evaluation of thermodynamic derivatives}
\label{subsec:methodology}

Based on the trajectories obtained in canonical ensemble, we compute Fourier components of species density using Eq.\eqref{eq:fourier_coeff_com}, at wavevectors compatible with the simulation box of size $(Lx,L_y,Lz)$ such that $\bm{q}=2\pi(\frac{n_x}{L_x},\frac{n_y}{L_y},\frac{n_z}{L_z})$ and $n_{x,y,z} \in \mathbb{Z}$ excluding $\bm{q}=\bm{0}$. $(n_x,n_y,n_z)$ triplets are chosen to span $\vert\bm{q}\vert$ vectors magnitude ranging from the minimum accessible values up to \qty{8}{\per\nano\m}.
For molecular system, \citealt{nichols2009pre} suggest to determine the fourier components of molecular density using the molecular center of mass of molecules with an isotropic form factor term based on the radius of gyration of the molecule to mimic an isotropic spatial distribution of molecular mass in real space.
In our work, when dealing with molecules we use a slightly modified definition of fourier components of molecular microscopic density:
\begin{equation}
  \rho_{i}(\bm{q}) = \sum_{i}\sum_{\alpha \in i} \frac{m_{\alpha}}{m_{i}} \exp(-\imath \bm{q}\bm{r_{\alpha}})
  \label{eq:fourier_coeff}
\end{equation}
Where \(m_{i}\) is the total mass of a molecule of species $i$, \(r_{\alpha}\) the position of an \(\alpha\) atom belonging to an \(i\) molecule. Thus the microscopic density spectrum for a given species \(i\) is determined through the total information given by all the position of the atoms constituting a molar species with a simple fractional mass weighting. This approach takes straightforwardly into account the mass spatial distribution of a given molecule in real space and it hence becomes useless to introduce a posteriori an isotropic $q$ dependant form factor. We believe that this definition results in a more accurate representation of the spatial distribution of the molecular mass and thus gives a better description of the wavevector dependency of the Fourier components of microscopic density through an averaging over all molecular orientations available in simulations.

The partial structure factors between species were averaged over wavevectors of same magnitude $q$ using Eq.\eqref{eq:sqadim}. For each $q$ value, we compute the mean $S_{ij}(q)$ values and the covariance $Cov(S_{ij}(q))$ matrix between the partial structure factors values over all the possible configurations.

Once obtained the mean values and covariance matrix of partial structure factor, we determine the fourier components of direct correlation function matrix $C(q)$ by inverting the $S(q)$ matrix for each (averaged) $q$ value using Eq.~\eqref{eq:directcf}.

Since the relationship between $\bm{C(q)}$ and $\bm{S(q)}$ matrices involves a Moore-Penrose pseudo-inversion, propagating uncertainties from $\bm{S(q)}$ to $\bm{C(q)}$ is an ill defined problem. To estimate uncertainties in $\bm{C(q)}$, we use a bootstrap method based on the mean values and covariances of $\bm{S(q)}$, assuming a multivariate Gaussian distribution of $\bm{S(q)}$ matrix. We randomly sample matrices $\bm{\tilde{S}(q)}$ and compute the corresponding $\bm{\tilde{C}(q)}$ matrix using Eq.~\eqref{eq:directcf}. This process is repeated until statistical convergence is achieved, at which point the standard deviation of elements of $\bm{\tilde{C}(q)}$ provides the uncertainties estimates.

This procedure is also used to determine the uncertainties of other derived observables, such as $\bm{A(q)}$, $\bm{D(q)}$, $\kappa_T(q)$, and $\bm{v(q)}$. From the obtained set of $\bm{C(q)}$ matrices and uncertainties, we extract the q-dependence of each element $C_{ij}(q)$ corresponding to the fourier components of direct pair correlations between species $i$ and $j$, and fit them independently using a least-squares method to the following low-q expansion:
\begin{equation}
C_{ij}(q)=C_{ij}^{(0)}+C_{ij}^{(2)} q^2+C_{ij}^{(3)} q^3 + C_{ij}^{(4)} q^4
\label{eq:cqexpansion}
\end{equation}

The fitted coefficients serve as initial guesses for a Markov Chain Monte Carlo (MCMC) sampling procedure based on Maximum Likelihood estimation \cite{foreman-mackeyEmceeMCMCHammer2013} applied to the full set of experimental $C(q)$ values. This sampling yields the distributions of the asymptotic expansion coefficients $C_{ij}^{(n)}$, from which we compute the mean values and covariance matrix of the extrapolated $C^{(0)}$ matrix, corresponding to the thermodynamic limit values. From the $C^{(0)}$ matrix and covariance, we directly determine extrapolated values and uncertainties for $\kappa_T$, $\bm{v}$, $\bm{D}$, and $\bm{\Gamma}$ via a bootstrap procedure analogous to the one described above.

\subsection{Determination of Gibbs enthalpy of excess}
\label{subsec:ivp}
Starting from equation \eqref{eq:chempotmole},in a multicomponent system the Gibbs free energy per molecule $G$ is:
\[
G = \sum_i x_i \mu_i(P, T, x_i),
\]
The Gibbs energy of mixing per molecule $G_{\text{mix}}$ is defined as:
\[
G_{\text{mix}} = G-\sum_i x_i\mu_i^0(P,T) = kT \sum_i x_i \ln \left( x_i\gamma_i\right)=H_{\text{mix}} - TS_{\text{mix}}
\]
where \( \gamma_i \) is the activity coefficient of species \( i \), the enthalpy of mixing per particle is given by $H_{\text{mix}}=H-\sum_{i}x_i H_{i}^{0}$ (where $H$ is the enthalpy per molecule of the mixture and $H_i^{0}$ the enthalpy per molecule of the pure phases at same $T$ and $P$) and $S_{\text{mix}}=S-\sum_i x_i S_i^0$ is the entropy of mixing per molecule.
The Gibbs energy of mixing per particle can be separated into both Gibbs energy of excess per molecule \( G_{\text{ex}} \) and the ideal Gibbs energy of mixture per molecule \( G_{\text{id}} \). Specifically,

\begin{equation}
G_{\text{ex}} = G_{\text{mix}} - G_{\text{id}} = kT \sum_i x_i \ln \left( \gamma_i \right) = H_{mix} - TS_{ex}
\label{eq:gexdef}
\end{equation}

and

\[
G_{\text{id}} =  kT \sum_i x_i \ln(x_i) = -T S_{\text{id}},
\]
where \( S_{\text{id}} \) is the ideal entropy of mixing per molecule. 
Although a general relation exists between the second derivatives of the Gibbs energy of mixing with respect to molar fraction (Hessian matrix) and the \( \Gamma \) matrix, as discussed in references \cite{van-brunt2023ea,monroe2015iecr,shimizu2018pccp}, this formalism is unnecessary for a binary mixture where the thermodynamic factor \(\Gamma\) is reduced to a single scalar related to the constrained second derivatives of the Gibbs energy of excess per molecule with respect to the molar fraction of one of the components\cite{ben-naim2006,taylor1991cec}:

\begin{eqnarray}
\beta \left( \frac{\partial^2 G_{\text{ex}}}{\partial x_1^2} \right)_{T,P,\Sigma} &=&\frac{ (\Gamma - 1)}{x_1 x_2} = \frac{1}{x_2} \left( \frac{\partial \ln(\gamma_1)}{\partial x_1} \right)_{T,P,\Sigma} \nonumber \\
&=&\frac{1}{x_1} \left( \frac{\partial \ln(\gamma_2)}{\partial x_2} \right)_{T,P,\Sigma}
\label{eq:gexsecderdef}
\end{eqnarray}

It is also possible to connect the activity coefficients of both species to the first derivative of $G_{\text{ex}}$\cite{taylor1991cec,ben-naim2006}:

\begin{equation}
\beta \left( \frac{\partial G_{\text{ex}}}{\partial x_1} \right)_{T,P,\Sigma} = \ln(\gamma_1) - \ln(\gamma_2)
\label{eq:gexderdef}
\end{equation}

The activity coefficients of each species can then be extracted combining relations \eqref{eq:gexdef} and \eqref{eq:gexderdef}.

To determine the values of \( \beta G_{\text{ex}}(x_1) \) and first derivatives from the thermodynamic factor, we use an iterative shooting method procedure similar to those described in references \onlinecite{galata2018fpe,petris2019jpcb}. Starting from defined boundary values of the primitive ($\beta G_{\text{ex}}(x_1=0)=\beta G_{\text{ex}}(x_1=1)=0$) and first derivative ($\ln(\gamma_1(x_1=1))=\ln(\gamma_2(x_1=0))=0$) corresponding to pure phase and using the second derivative values at various concentrations between those endpoints, it is possible by successive integration to determine the first derivative and primitive $\beta G_{\text{ex}}$ for the given concentrations points. To determine the first derivative, we must use as an initial guess the activity coefficient of one species at infinite dilution \(\gamma_1(x_1=0)=\gamma_{1,\infty}, \gamma_2(x_1=1)=\gamma_{2,\infty}\). Starting from one endpoint we calculate the firsts derivatives and primitive at intermediate molar concentration up to the other endpoint using an Adams-Moulton scheme. The values of activity coefficient and $G_{ex}$ hence obtained at the final endpoint are then compared with the aforementioned extremum conditions and the difference between final values are used in a minimization procedure to modify our starting values of activity coefficients at infinite dilution. During the minimization procedure, we integrate separately the functions both upwards and backwards and minimize simultaneously on both sides.
Finally, the uncertainties on \( G_{ex}(x_1) \) and the activity coefficients are obtained through a bootstrap procedure, by repeating the above minimization process while using uncorrelated random variables \( \tilde{\Gamma}(x_1) \) associated with Gaussian distributions, where the mean values and standard deviations correspond to the values of \( \Gamma(x_1) \) and their uncertainties.

\section{Results}
\subsection{Binary Lennard-Jones mixture}
\begin{figure*}
  \includegraphics[width=\textwidth]{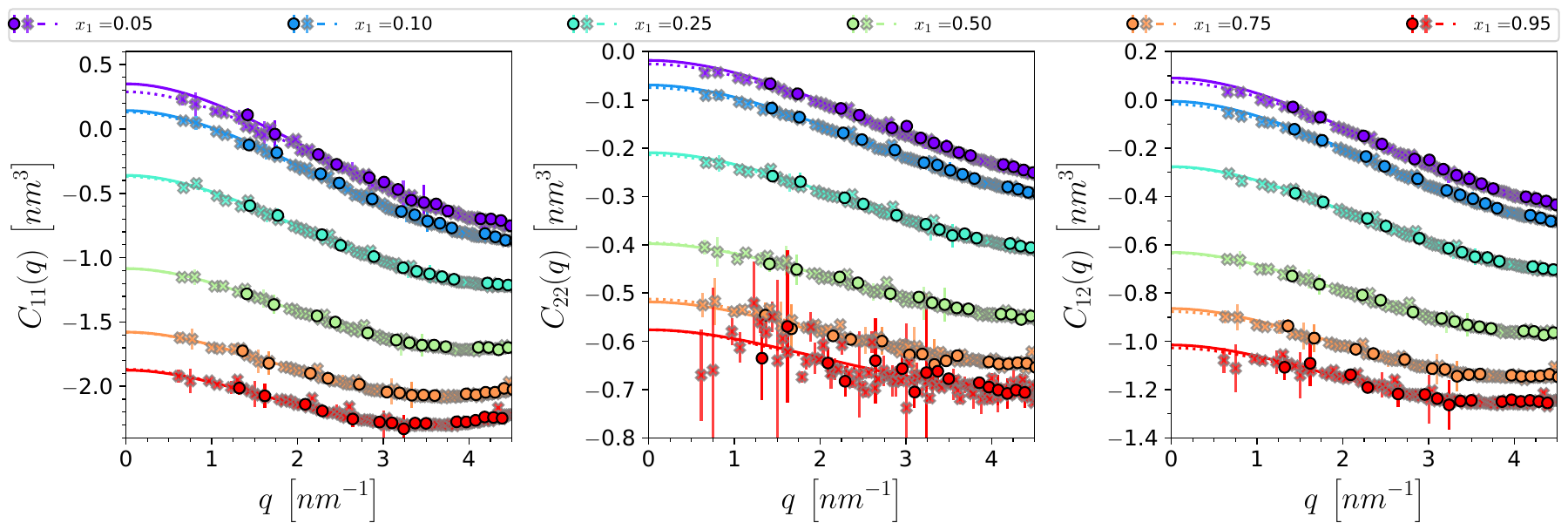}
  \includegraphics[width=\textwidth]{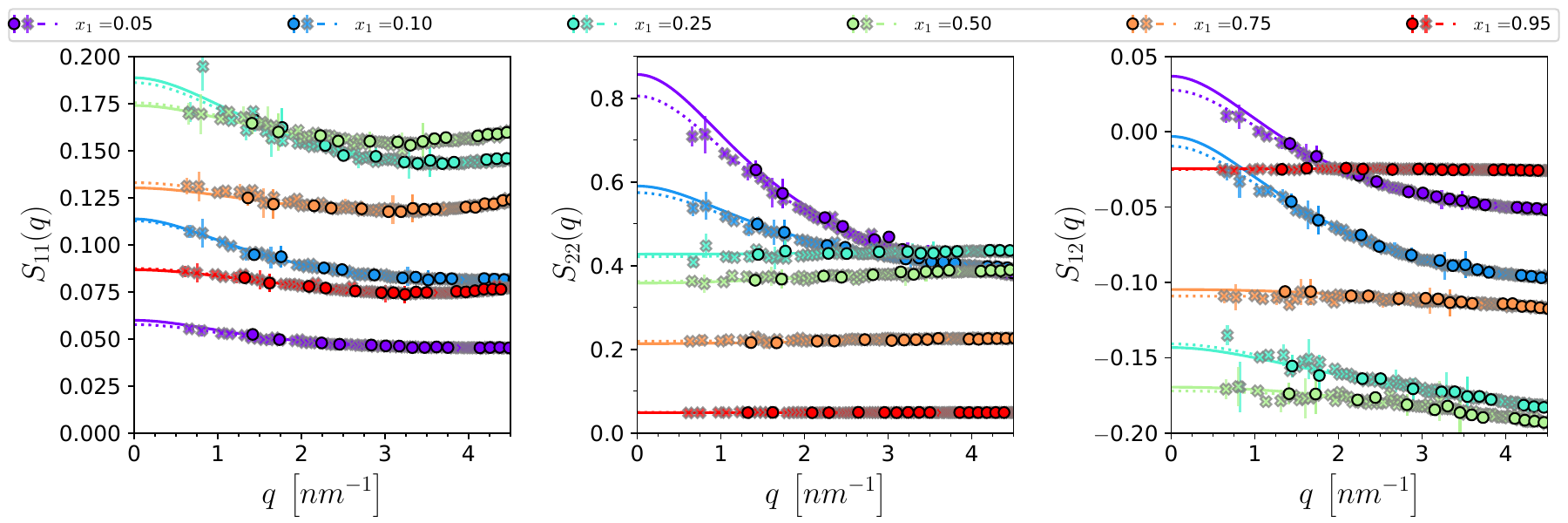}
  \caption{
    Top: Fourier components of the direct pair correlation function. Left: $C_{11}(q)$, middle: $C_{22}(q)$, right: $C_{12}(q)$.\\
    Bottom: Partial structure factors. Left: $S_{11}(q)$, middle: $S_{22}(q)$, right: $S_{12}(q)$.\\
    Circles: data for a system with $N=2000$ particles; Crosses: data for a system with $N=20000$ particles; Solid lines: fit results for the $N=2000$ systems; Dotted lines: fit results for the $N=20000$ systems.\\
    Purple: $x_1=0.05$, Blue: $x_1=0.1$, Cyan: $x_1=0.25$, Green: $x_1=0.5$, Orange: $x_1=0.9$, Red: $x_1=0.95$.
  }
  \label{fig:sqcqbinary}
\end{figure*}

We begin by investigating a binary Lennard-Jones mixture previously studied by \textcite{galata2018fpe} using methods based on corrected radial distribution functions and spatial block analysis. We adopt the same potential parameters as in the original study for the real mixture: \(\epsilon_1=\SI{3.1346}{\kilo\joule\per\mol}, \sigma_1=\SI{4.95}{\angstrom}\), and \(\epsilon_2=\SI{1.8706}{\kilo\joule\per\mol}, \sigma_2=\SI{3.8}{\angstrom}\), with Lorentz-Berthelot mixing rules: \(\sigma_{ij}=\frac{\sigma_i + \sigma_j}{2}, \epsilon_{ij}=\sqrt{\epsilon_i\epsilon_j}\).

We simulate 21 binary Lennard-Jones mixtures with mole fractions ranging from \(x_1=0\) to \(x_1=1\) in steps of 0.05, using \(N=2000\) particles, significantly fewer than the 20000 particles used in the original study. To assess finite-size effects, we additionally simulate seven systems with \(N=20000\) particles at selected mole fractions (\(x_1 = 0.05, 0.1, 0.25, 0.5, 0.75, 0.9, 0.95\)). All simulations follow the methodology described in Section~\ref{subsec:methodology}, conducted at \(T=\SI{350}{\kelvin}\) and \(P=\SI{200}{\atmosphere}\).

For each dataset, we compute \(S(q)\) and \(C(q)\) following the procedure in Section~\ref{subsec:methodology}. Looking at the results for \(C_{ij}(q)\) in Fig.~\ref{fig:sqcqbinary}, we observe excellent agreement between data and fitted curves for both system sizes at identical compositions. No systematic deviations are present, suggesting negligible size effects. Minor discrepancies in \(C_{11}(q)\) at \(x_1=0.05\) and in \(C_{22}(q)\) at \(x_1=0.95\) are attributed to statistical noise, as the smaller systems contain only 100 particles of one species. Except for these extreme cases, the extrapolated \(C^{(0)}_{ij}\) values at the thermodynamic limit differ by less than 5\% between the large and small systems.

\begin{figure}
  \includegraphics[width=\columnwidth]{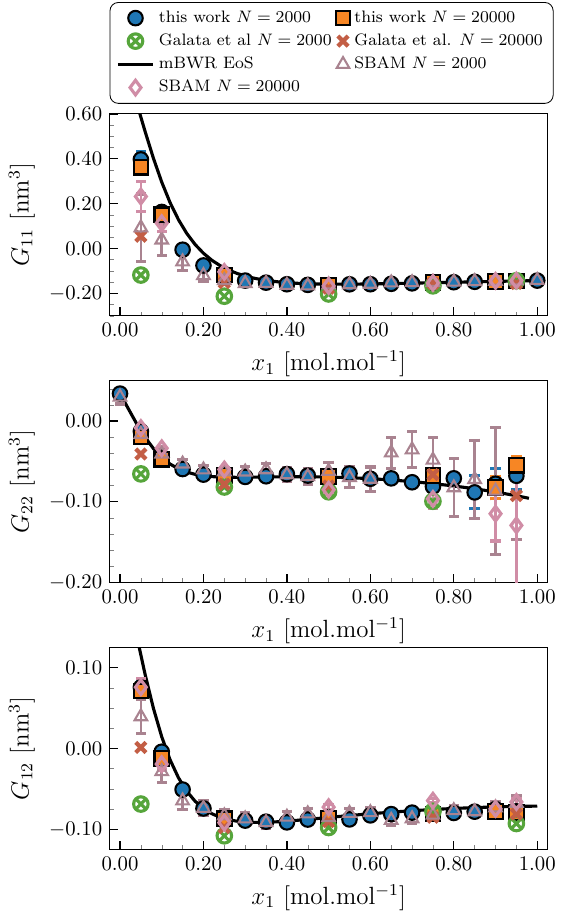}
  \caption{Extrapolated Kirkwood–Buff integrals versus mole fraction of species 1. Blue circles and orange squares: our results for 2000 and 20000 particles, respectively. Green crossed circles and purple crosses: results from Ref.~\citenum{galata2018fpe} for 2000 and 20000 particles, respectively using SBAM method. Brown triangle and purple diamond: SBAM method on our data respectively $N=2000$ and $N=20000$. Black lines: values computed using the mBWR/vdW1 EoS~\cite{mayRiemannianGeometryStudy2012,mayErratumRiemannianGeometry2012,johnsonLennardJonesEquationState1993}.}
  \label{fig:gijbinary}
\end{figure}

By transforming the extrapolated \(C^{(0)}_{ij}\), we derive thermodynamic properties including the isothermal compressibility, partial molar volumes, Kirkwood–Buff integrals, chemical potential derivatives, and the thermodynamic factor. These are validated against both the published data from \cite{galata2018fpe} and predictions from the mBWR EoS using the van der Waals one-fluid (vdW1) model (see Supplementary Material).

Since \(C^{(0)}_{ij}\) values are nearly identical between system sizes, the resulting Kirkwood–Buff integrals (KBI) are effectively size-independent, as shown in Fig.~\ref{fig:gijbinary}. Small differences observed at very low and high mole fractions reflect extrapolation uncertainties, yet remain within error margins and show no systematic bias.

The KBI values closely match those calculated from the mBWR/vdW1 EoS. Although for low \(x_1\), small differences appear, but the sharp decline in \(G_{11}\) and \(G_{22}\) is captured well. The agreement in \(G_{12}\) is nearly perfect.
We compare the KBI values obtained by our method to the ones obtained using spatial block analysis method (SBAM) method presented in \cite{galata2018fpe} and calculated on our samples. Although deviations in $G_{11}$ at low \(x_1\) suggest an important underestimation of KBI values found using SBAM methods. Agreement improves significantly at higher \(x_1\), where all methods converge. We note that the KBI reported for smaller system by \cite{galata2018fpe} are systematically far of the other data.
Concerning $G_{22}$ our SBAM results show an increase in the dispersity of the results ,especially in the smaller system, for $x_1>0.6$. Deviations from EoS predictions are increased in comparison of the $G_{22}$ values found following our method.
Finally, concerning $G_{12}$, despite a small underestimation of EoS prediction by the SBAM method, we underline thant no differences are found comparing the values obtained by our method and the SBAM method on the greater system.
We stress that the overall agreement between SBAM and reciprocal-space analyses is reasonable within uncertainties, particularly at intermediate compositions.
Because KBIs are directly related to isothermal compressibility and partial molar volumes, we also compare our compressibility values derived from KBIs to those obtained from NPT volume fluctuations and EoS predictions. Partial molar volumes and excess volumes are similarly compared (see Supplementary Material).
\begin{figure}
  \includegraphics[width=\columnwidth]{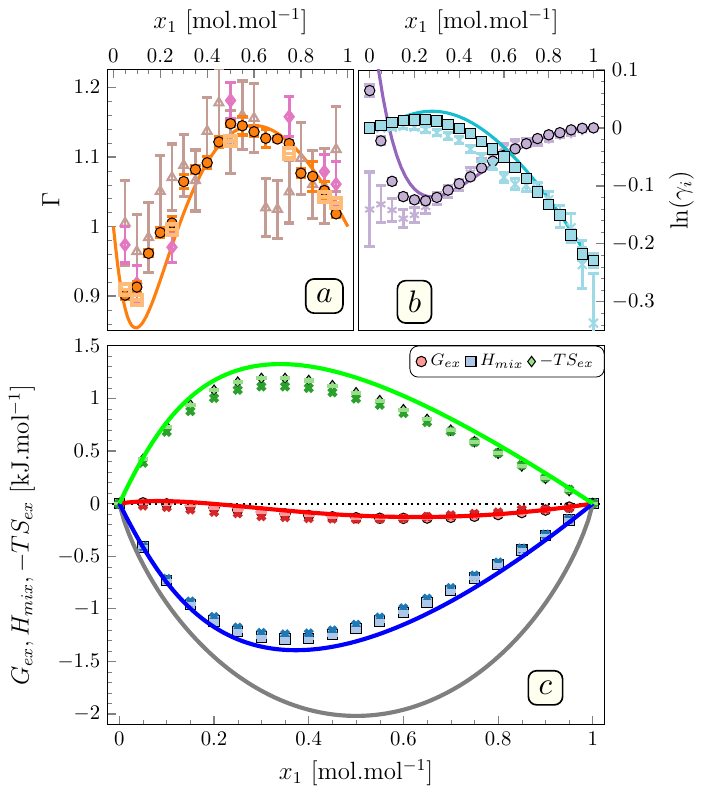}
    \caption{
  a) Thermodynamic factor \(\Gamma\) versus mole fraction \(x_1\). Orange circles: values from reciprocal space method $N=2000$; light orange square: values from reciprocal space method $N=20000$; Brown triangles: SBAM method $N=2000$; Purple diamonds: SBAM method $N=20000$; orange line: mBWR/vdW1 EoS prediction~\cite{mayRiemannianGeometryStudy2012,mayErratumRiemannianGeometry2012,johnsonLennardJonesEquationState1993}.\\
  b) Logarithm of activity coefficients on $N=2000$ samples. Purple circles: species 1/ reciprocal space method; cyan squares: species 2/reciprocal space method; Purple stars: species 1/SBAM Method; Cyan crosses: species 2/SBAM method; lines: EoS predictions.\\
  c) Excess thermodynamic properties versus \(x_1\) from $N=2000$ samples: excess enthalpy (blue squares), excess Gibbs free energy (red circles), \(-T S_{\text{ex}}\) (green diamonds). Data from \cite{galata2018fpe} and EoS predictions are included as crosses and lines respectively. Gibbs energy of ideal mixing is also reported (gray dashed line).}
  \label{fig:thermobinary}
\end{figure}
The thermodynamic factor \(\Gamma\) values, obtained using our method and SBAM, are shown in Fig.~\ref{fig:thermobinary}-a as a function of \(x_1\) and compared with predictions from the mBWR/vdW1 EoS. Since \(\Gamma\) is related to the Hessian matrix of the Gibbs free energy of mixing, \(G_{\mathrm{mix}}\), positive values are expected for a stable mixture~\cite{taylor1991cec,ben-naim2006}. For an ideal mixture, \(\Gamma = 1\), and deviations from unity indicate non-ideal behavior.

First, $\Gamma$ values obtained with reciprocal space method on sample of $2000$ and $20000$ atoms does not display significant differences, an observation in agreement with the closely related values found in KBI's. A comparison of our reciprocal space method values with EoS predictions shows good agreement for \(x_1 > 0.3\), where \(\Gamma\) crosses unity, increases up to \(x_1 = 0.55\), and then gradually decreases back toward one. For \(x_1 < 0.3\), the EoS predicts a distinct minimum around \(x_1 = 0.1\), with \(\Gamma\) returning to the ideal value of one as \(x_1 \to 0\). In contrast, our results show no clear minimum in this region, suggesting that the minimum may lie between the pure component 2 phase and \(x_1 = 0.05\). Despite the absence of a visible minimum, our computed values of \(\Gamma\) remain close to those predicted by the EoS and are slightly less than one, indicating that both approaches capture the same qualitative behavior: at low \(x_1\), \(\Gamma < 1\) reflects reduced miscibility compared to the ideal case, while for \(x_1 > 0.3\), the mixture becomes more miscible than an ideal solution.
$\Gamma$ values were also computed using SBAM method on both system size, if the results show the same overall global trend than EoS predictions, values denote an increased dispersity with larger error bars than our proposed method. More problematic the data show a systematic shift at low $x_1$ molar fraction for the smaller system, a direct consequence of the underestimation of $G_{11}$ in this range.

Using the iterative shooting method described in Section~\ref{subsec:ivp}, we also extracted the logarithm of the activity coefficients (Fig.~\ref{fig:thermobinary}-b) from data analyzed with our reciprocal space method or SBAM method and excess thermodynamic properties (Fig.~\ref{fig:thermobinary}-c). Looking to the values for species 1 using our proposed method, \(\ln(\gamma_1)\) is positive at infinite dilution, then rapidly decreases to a negative minimum near \(x_1 = 0.2\), before gradually returning toward zero. This trend is consistent with EoS predictions, although the values obtained at infinite dilution and low \(x_1\) are somewhat smaller than those predicted by the EoS. A positive \(\ln(\gamma_1)\) at low \(x_1\) indicates that species 1 prefers self-association over mixing, whereas the negative values appearing near \(x_1 = 0.2\) indicate a higher affinity between species 1 and 2 than in an ideal solution. When analyzed with SBAM method, the trend of \(\ln(\gamma_1)\) match the values of our proposed method and EoS predictions when $x_1>0.3$, we note however, a totally different picture at low $x_1$ where \(\ln(\gamma_1)\) remains negative, a behaviour in opposition to the EoS predictions and the values obtained with reciprocal space method but consistent with the $\Gamma\approx 1$ values found at low $x_1$ using SBAM method.

For species 2, the values of \(\ln(\gamma_2)\) obtained using our method increases slightly at low \(x_1\), peaks around \(x_1 = 0.2\), then decreases and becomes negative near \(x_1 = 0.4\). At higher concentrations of \(x_1\), it decreases nearly linearly toward a negative value at infinite dilution. This behavior aligns well with the EoS predictions and resembles the trends reported in Ref.~\cite{fingerhut2019fpe} for the "LJ4" system. The slight increase in \(\ln(\gamma_2)\) at low \(x_1\) suggests marginally increased self-affinity for species 2 compared to the ideal case, followed by a stronger decrease indicating enhanced mixing with species 1 at higher concentrations.
When determined with SBAM method, \(\ln(\gamma_2)\) does not display the small increase around 0.2 and values between $0.1$ and $0.5$ are significantly decreased in comparison to EoS predictions but realign for $x_1>0.6$.

When both \(\ln(\gamma_1)\) and \(\ln(\gamma_2)\) are negative, the excess Gibbs free energy of mixing is necessarily negative, indicating improved miscibility relative to the ideal mixture, a finding consistent with the behavior of \(\Gamma\) in this concentration regime. 
When analyzing the system with our proposed method we show that towards low \(x_1\), a subtle interplay between the increasing affinity among species 2 and decreasing self-affinity of species 1, modulated by species composition, leads to a transition from lower to higher miscibility near \(x_1 = 0.2\), even though the activity coefficients do not change sign, we must underline that when analyzing our data using SBAM method, this level of discussion can not be reached.

The excess Gibbs free energy \(G_{\text{ex}}\), enthalpy of mixing \(H_{\text{mix}}\), and the energetic counterpart of the excess entropy of mixing, \(-T S_{\text{ex}}\), are shown in Fig.~\ref{fig:thermobinary}-c ,for the results obtained with our proposed method alone, along with the EoS predictions and the data from Ref.~\cite{galata2018fpe}. At first glance, the agreement between the datasets is very good. While the EoS results show slight deviations in the magnitudes of \(H_{\text{mix}}\) and \(-T S_{\text{ex}}\), the overall trends and shapes of the curves are preserved.

The values of \(G_{\text{ex}}\) are found to be an order of magnitude smaller than those of the ideal Gibbs free energy of mixing \(G_{\text{id}}\), clearly indicating full miscibility of the system. We observe negative \(H_{\text{mix}}\) values, consistent with an exothermic mixing process, with a minimum located around \(x_1 = 0.4\). The relatively large magnitude of \(H_{\text{mix}}\), when compared to \(G_{\text{ex}}\), results in correspondingly large values of \(-T S_{\text{ex}}\) with opposite sign, indicating a significant excess entropic contribution opposing the enthalpic gain.

At low values of \(x_1\), slightly positive \(G_{\text{ex}}\) values are observed, in qualitative agreement with the EoS predictions. This observation is consistent with the thermodynamic factor \(\Gamma < 1\) in the same composition range, and with the opposite signs of the logarithms of the activity coefficients of the two species. Notably, such positive \(G_{\text{ex}}\) values are absent in the data of Ref.~\cite{galata2018fpe}. The authors attribute this discrepancy to the limitations of the one-fluid model in reproducing the low-composition behavior of the phase diagram. 

We partially disagree with this explanation and note that the thermodynamic quantities computed using our proposed method, while not perfectly reproduced by the mBWR/vdW1 EoS at low $x_1$, show significantly better agreement with the EoS predictions than those reported in Ref.~\cite{galata2018fpe}. This improved consistency is strongly supported by our KBIs, partial molar volumes, and isothermal compressibilities, when obtained via our reciprocal-space extrapolation methodology.

In conclusion, when computed using the SBAM approach, the KBIs and associated thermodynamic derivatives are consistent with those reported in Ref.~\cite{galata2018fpe}, but show larger deviations from the EoS predictions. In contrast, our reciprocal-space extrapolation methodology provides results that remain closer to the EoS across the entire composition range, including at the phase diagram endpoints where species concentrations are very low.
%Lastly, it is important to emphasize that Ref.~\cite{galata2018fpe} explicitly reported that the RDF-based method failed to converge reliably for systems comprising only 2,000 particles, particularly at extremal concentrations. %In contrast, our reciprocal-space method yields well-converged KBI values even for small systems and across all mole fractions, a level of precision that we were unable to achieve using the SBAM method on the same samples.
%In contrast, our reciprocal-space method yields well-converged KBI values even for small systems and across all mole fractions, providing a level of numerical precision and thermodynamic consistency that was not attained with SBAM on the same samples.
Lastly, it is important to emphasize that Ref~\cite{galata2018fpe} explicitly reported that the RDF-based method failed to converge reliably for systems comprising only 2,000 particles, particularly at extremal concentrations. By contrast, when compared with SBAM on the same trajectories our reciprocal-space method yields well-converged KBI values even for small systems and across all mole fractions, providing a level of numerical precision and thermodynamic consistency that was not attained with SBAM on the same samples.

\subsection{Quaternary Lennard-Jones mixture}

\begin{figure}
  \includegraphics[width=\columnwidth]{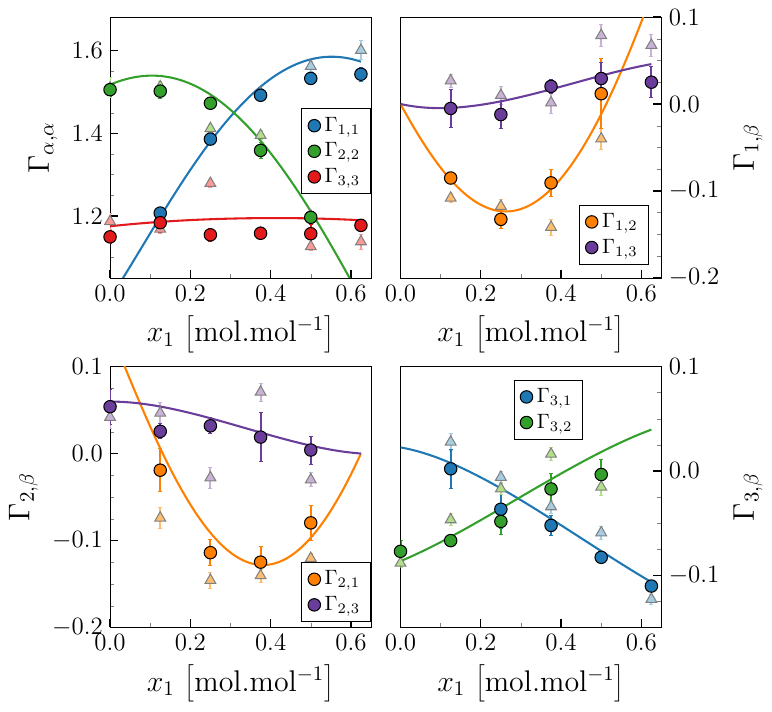}
  \caption{Thermodynamic factors obtained using the relationship \eqref{eq:gamma_matrix} for the quaternary system, with species 4 chosen to maintain the constraint during differentiation. Circles: this work; Triangles: ref. \citenum{fingerhut2020mp}, Lines: mBWR/vdW1 EoS.}
  \label{fig:quatgamma}
\end{figure}

We base our Kirkwood-Buff inversion approach on the elegant formalism introduced by \textcite{oconnell1975jsc}, which establishes a linear algebra relationship between the KBIs matrix and the \(\Gamma\) matrix of thermodynamic factors. As previously discussed, in binary systems, the \(\Gamma\) matrix reduces to a single scalar, making the connection to the three independent KBIs values straightforward, and the use of matrix formalism optional.

In ternary mixtures, however, one must relate four independent elements of the \(\Gamma\) matrix to six independent KBIs values. Without linear algebra formalism, deriving these relationships requires laborious and error-prone algebraic manipulations \cite{ruckensteinEntrainerEffectSupercritical2001}. For quaternary systems, where ten independents KBIs values must be related to nine independent elements of the \(\Gamma\) matrix, establishing these connections becomes a genuine "tour de force" \cite{fingerhut2020mp}.

Since these relationships have already been developed and successfully applied to compute thermodynamic derivatives in Lennard-Jones quaternary mixtures under liquid-like supercritical conditions, we adopt this system and apply the methodology \ref{subsec:methodology} accordingly to carry out the analysis within the linear algebra framework.

The simulations by \textcite{fingerhut2020mp} employed Lennard-Jones potentials with \(\sigma_1 = \sigma_2 = \sigma_3 = \sigma_4\), \(\epsilon_2 = \frac{5}{6} \epsilon_1\), \(\epsilon_3 = \frac{25}{36} \epsilon_1\), \(\epsilon_4 = \frac{125}{216} \epsilon_1\), and a modified Lorentz-Berthelot mixing rule: \(\epsilon_{ij} = \zeta \sqrt{\epsilon_i \epsilon_j}\) with \(\zeta = 1.5\) when $i\neq j$ and 1 otherwise. These mixtures were studied at a reduced pressure \(P^* = P \sigma_1^3 / \epsilon_1 = 4\) and reduced temperature \(T^* = kT / \epsilon_1 = 2.2\).

We reproduce the same simulations in a non-reduced unit system by selecting \(\sigma_1 = \qty{5}{\angstrom}\) and \(\epsilon_1 = \qty{2}{\kilo\J\per\mol}\), which corresponds to a state point located at \(T = \qty{529}{\K}\) and \(P = \qty{1062}{\bar}\) on the phase diagram. We simulate 8000 particles for a subset of the quaternary phase diagram, fixing the mole fractions of species 3 and 4 at \(x_3 = \frac{1}{8}\) and \(x_4 = \frac{1}{4}\), and exploring six state points evenly spaced along the \((x_1, x_2 = \frac{5}{8} - x_1)\) line from \(x_1 = 0\) to \(x_1 = \frac{5}{8}\).

Configurations and subsequent thermodynamic analyses are performed according to the protocol described in Section~\ref{subsec:methodology}. In a quaternary system, the matrix of thermodynamic factors contains \(3 \times 3\) independants derivatives of chemical potentials. For consistency with \cite{fingerhut2020mp}, we also select species 4 to impose the mole fraction constraint when constructing the \(\Gamma\) matrix. We also compute the thermodynamic properties and derivative of the systems using the mBWR/vdW1 EoS for the quaternary mixture. In Supplementary Material, we report isothermal compressibility and partial volumes of species determined via KB inversion and mBWR/vdW1 EoS. We first want to underline the surprisingly good agreements between our values and the EoS, indicating the unexpected reliability of mBWR/vdW1 EoS for quaternary LJ mixture.

The resulting thermodynamic factor coefficients are shown in Fig.~\ref{fig:quatgamma} as a function of \(x_1\), with results from \textcite{fingerhut2020mp} and mBWR/vdW1 EoS results superimposed for comparison.

Focusing on the diagonal elements \(\Gamma_{\alpha,\alpha}\) of thermodynamic matrix, we observe close agreement between experimental data sets for species 1 and 2, those thermodynamic factors are also in really fair agreement with EoS. Our results for species 3 conserve the quasi flat behavior as the EoS results, but are slightly shifted to smaller values. For the cross-derivative terms, our values are generally in close agreement with the EoS values
Comparing with results of \citenum{fingerhut2020mp} we sometimes observe systematic shifts (e.g., \(\Gamma_{2,1}\), \(\Gamma_{3,2}\)), changes in the location of extrema (\(\Gamma_{1,2}\), \(\Gamma_{1,3}\)), or even different overall trends (\(\Gamma_{2,3}\)). Despite these small discrepancies, the comparison remain satisfying within the error bars reported in both studies. We want to underline that our results are systematically closer to the EoS values than those of \cite{fingerhut2020mp}.
The good agreement between our values and EoS results underscores the reliability of our method for extracting thermodynamic information in multicomponent systems, far beyond the binary case. Furthermore, the matrix formalism provides a direct and rigorous way to obtain derivatives of chemical potentials with respect to species composition,a necessary foundation for studying Maxwell-Stefan diffusion in complex systems.

\subsection{Hexane-Ethanol mixture}
\begin{figure*}
  \includegraphics[width=\textwidth]{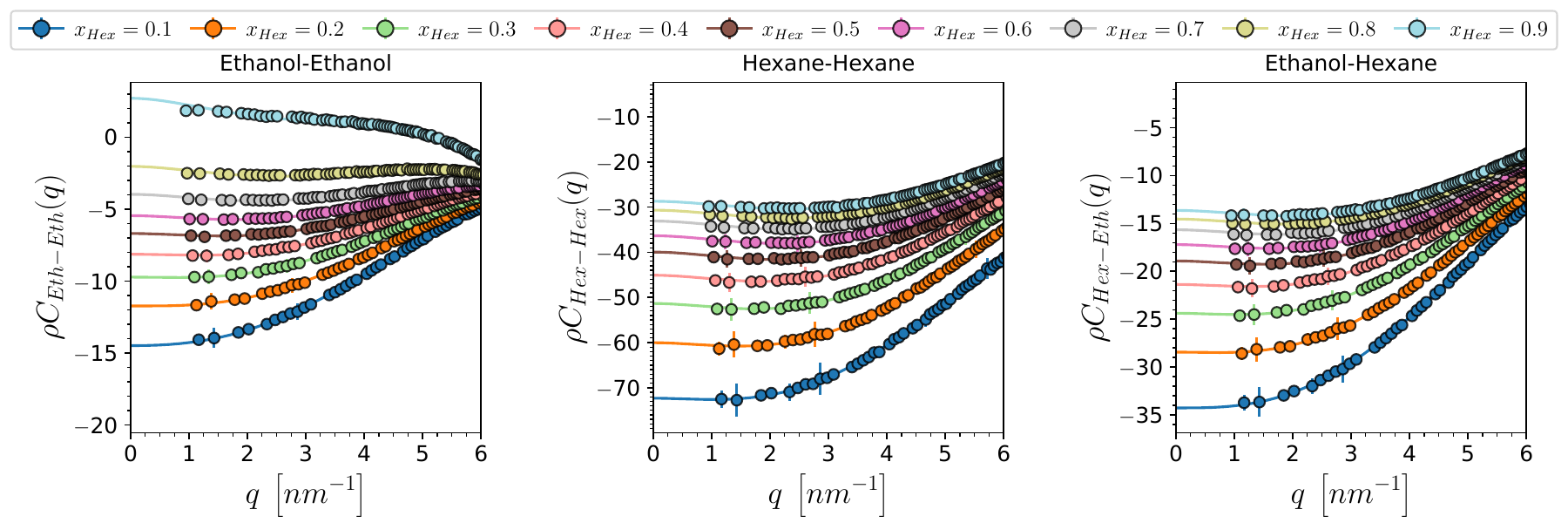}
  \includegraphics[width=\textwidth]{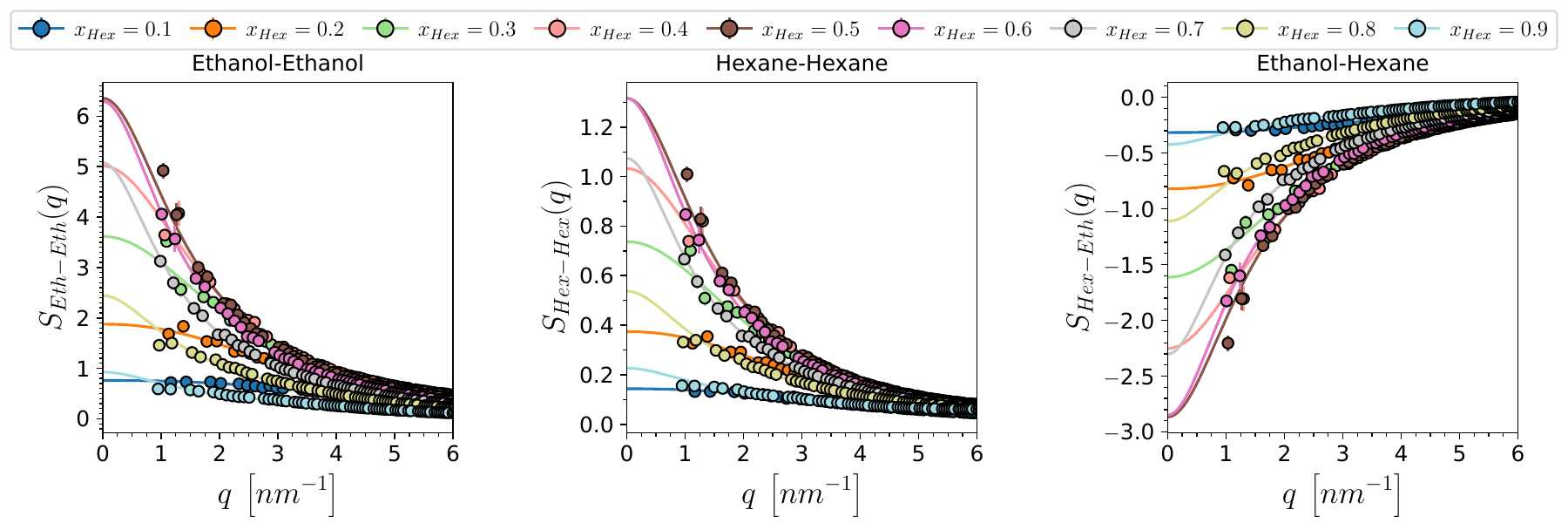}
  \caption{
  Top: Fourier components of direct pair correlation functions multiplied by particle density along $q$ for different hexane molar fractions. From left to right: ethanol–ethanol, hexane–hexane, and hexane–ethanol interactions.
  Bottom: Partial structure factors as a function of $q$ for different hexane molar fractions. From left to right: ethanol–ethanol, hexane–hexane, and hexane–ethanol interactions. \newline
  In both cases, circles represent simulation data, while solid lines correspond to fitted results based on the direct correlation function.}
  \label{fig:ethhex-sqcq}
\end{figure*}

\begin{figure}
  \includegraphics[width=\columnwidth]{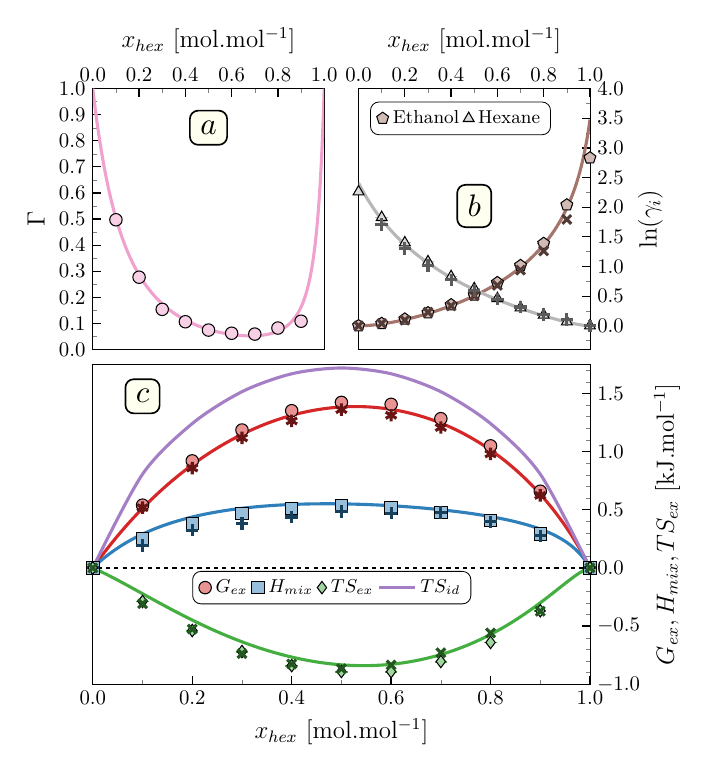}
  \caption{
  (a) Thermodynamic factor of the mixture as a function of hexane molar fraction. Circles denote values obtained from this work; solid line represents predictions from the Wilson model based on experimental data \cite{hongo1994jced}. \newline
  (b) Activity coefficients of ethanol (brown) and hexane (gray) in the binary mixture as a function of hexane molar fraction. Filled markers: this work; crosses and pluses: values from \cite{petris2019jpcb}; solid lines: Wilson model from experimental data \cite{hongo1994jced}. \newline
  (c) Excess Gibbs energy (red), enthalpy of mixing (blue), and excess entropy multiplied by temperature (green). Filled markers: this work; unfilled markers: data from \cite{petris2019jpcb}; solid lines: thermodynamic quantities from the Wilson model \cite{hongo1994jced}; purple line: ideal entropy of mixing multiplied by temperature.}
  \label{fig:gexhexeth}
\end{figure}

Hexane–ethanol mixtures are known to exhibit notable deviations from ideal mixture behavior due to the complex interplay between the apolar nature of hexane and the amphiphilic character of ethanol. This dual nature can give rise to microheterogeneities \cite{murdochInfraredSpectroscopyEthanol2002, pozarMicroheterogeneityClusteringBinary2016}, resulting in significant concentration fluctuations. The phase diagram of the mixture was previously determined using KB approach by \textcite{petris2019jpcb}, based on the TraPPE-UA force field \cite{chen2001jpcb} and a modified version of the spatial block analysis method \cite{cortes-huerto2016tjocp}, which accounts for complex molecular structure beyond the center-of-mass approximation.
Microheterogeneities in alkane–alcohol mixtures are often accompanied by large concentration fluctuations, which can influence mixture stability. For hexane–ethanol systems at ambient temperature and pressure, the experimentally observed thermodynamic factor is slightly positive for hexane molar fractions between 0.4 and 0.9. These values indicate proximity to critical behavior \cite{hongo1994jced, seo2024jpcb}, making this system a valuable test case for practical applications of the KB inversion formalism.
Although both components are neutral molecules, ethanol–ethanol interactions include a significant electrostatic component. Notably, the combination of dispersion and electrostatic potentials defined at the atomic level does not necessarily yield a $r^{-6}$ attractive tail at large distance in the effective molecular-level potential. As a consequence, the low-$q$ behavior of the molecular direct pair correlation function in Fourier space may include terms beyond the power-law expansion \eqref{eq:cqexpansion}.
We employ the same force field \cite{chen2001jpcb} as \textcite{petris2019jpcb} and conduct 11 simulations of 4000 molecules each, covering the full range of hexane molar fractions (0 to 1) at $T = \qty{298.15}{\kelvin}$ and $P = \qty{1}{\bar}$. The molecular form of Eq.~\ref{eq:fourier_coeff} is used to extract the density fluctuation spectrum. Fig.~\ref{fig:ethhex-sqcq} presents the partial structure factors and fourier components of direct correlation functions.
At a glance, the partial structure factors reveal a pronounced increase at low-$q$ for ethanol–ethanol and hexane–hexane structure factors. In the same $q$ region, a strong negative partial structure factor appears in ethanol–hexane structure factors. These features persist across all molar fractions but are particularly intense for hexane fractions between 0.3 and 0.7, indicating substantial local compositional fluctuations.
In contrast to the structure factors, the Fourier components of the direct pair correlation functions, \(C_{ij}(q)\), exhibit a monotonic and nearly constant behavior in the low-\(q\) region, where the structure factors increase sharply. For hexane molar fractions between 0.4 and 0.9, all \(C_{ij}(q)\) curves either reach a plateau or display only slight curvature at low \(q\), which facilitates reliable extrapolation using the low-\(q\) expansion (Eq.~\ref{eq:cqexpansion}) to determine \(C_{ij}^{(0)}\).
No modification to the asymptotic expansion of \(C_{ij}(q)\) appears necessary to account for changes in molecular interaction potentials arising from mean electrostatic effects. We hypothesize that the low-\(q\) components of the direct pair correlation functions are subject to long-range electrostatic screening, and that their average behavior can be described using the same power-law form typically applied to Lennard-Jones potentials.

Supplementary material includes comparisons of density, excess volume, isothermal compressibility, and KB integrals with available experimental data \cite{ormanoudisVolumetricPropertiesBinary1991, sauermann1995fpe, hongo1994jced, seo2024jpcb, jimenezSurfaceTensionsRefractive2000}. The agreement is generally good, although excess volume tends to be overestimated a known flaw of the Trappe-UA force field\cite{petris2019jpcb}. Nevertheless, the thermodynamic values derived from KB analysis remain consistent with experimental measurements \cite{seo2024jpcb}.

Figure~\ref{fig:gexhexeth}-a displays the thermodynamic factor as a function of hexane molar fraction, showing excellent agreement with predictions from the Wilson model based on experimental data \cite{hongo1994jced}. The thermodynamic factor decreases rapidly at low hexane fractions and increases significantly as the hexane content approaches unity. For intermediate fractions (0.3 to 0.9), the thermodynamic factor remains below 0.2, reaching a minimum of less than 0.1 at $x_\mathrm{hex} = 0.6$. These small positive values confirm the overall stability of the mixture despite pronounced fluctuations.

Activity coefficients (Figure~\ref{fig:gexhexeth}-b) and excess thermodynamic properties (Figure~\ref{fig:gexhexeth}-c) are computed using iterative shooting method (\ref{subsec:ivp}). The logarithms of activity coefficients are compared with reference values \cite{petris2019jpcb} and Wilson model predictions based on experimental data \cite{hongo1994jced}, three cases showing really close agreement. Logarithm of activity coefficients are positive for all compositions and both species, indicating significant deviation from ideal behavior and strong preferential interactions between like molecules.

This observation is further supported by the strongly positive excess Gibbs free energy and the notably high positive values of the enthalpy of mixing, as shown in Fig.~\ref{fig:gexhexeth}-c. The positive \(H_{\text{mix}}\) values indicate an endothermic mixing process. The excess entropy is significantly negative, reflecting a substantial reduction in mixing entropy relative to that of an ideal solution, evidence of a mixed state that is markedly more ordered than the ideal case. Despite the influence of these excess contributions, the overall mixing terms ensure that \(T S_{\text{mix}}\) remains positive and \(G_{\text{mix}}\) remains negative, confirming the thermodynamic stability of the system.

All excess thermodynamic properties are in good agreement with both prior simulation results~\cite{petris2019jpcb} and experimental measurements~\cite{hongo1994jced, seo2024jpcb}, demonstrating that the present Kirkwood-Buff (KB) inversion methodology accurately captures thermodynamic behavior in polar molecular systems characterized by strong fluctuations near the stability limit. As observed for binary Lennard-Jones mixtures, the required system size for obtaining reliable data is significantly reduced—by more than a factor of two—compared to standard techniques such as radial distribution function analysis or spatial block analysis models.

\subsection{Urea Water mixture}
\begin{figure}
  \includegraphics[width=\columnwidth]{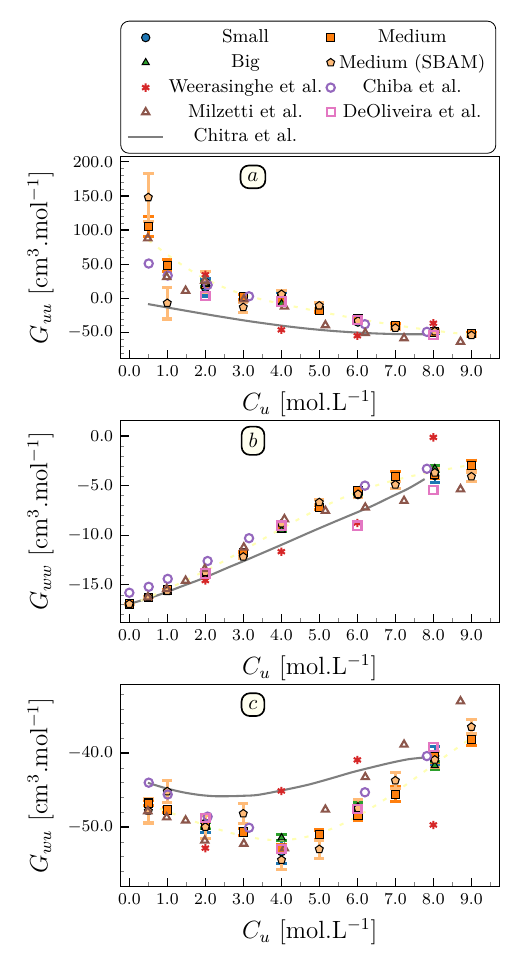}
  \caption{Kirkwood-Buff integrals as a function of urea molality. (a) Urea–Urea, (b) Water–Water, (c) Urea–Water. Filled markers indicate results from our simulations: small system (blue circles), medium system (orange squares), Big system (green triangles), Medium (SBAM) (light orange pentagones). Unfilled markers are data from previous studies: red stars \citenum{weerasinghe2003jpcb}, purple circles \citenum{chibaKirkwoodBuffIntegrals2016}, brown triangles \citenum{milzettiConvergenceKirkwoodBuff2018}, and pink squares \citenum{deoliveiraIBITargetingCumulative2016}. gray line represent experimental values from \citenum{chitraMolecularAssociationSolution2002}.}
  \label{fig:gijureawater}
\end{figure}
\begin{figure}
\includegraphics[width=\columnwidth]{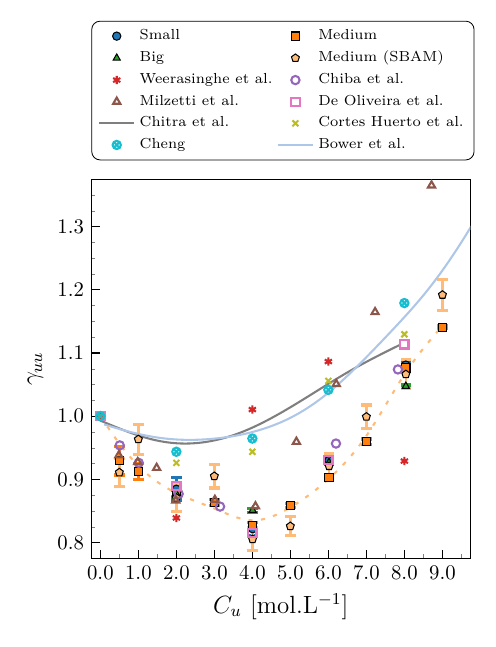}
\caption{Derivative of the urea activity coefficient (in molarity units) as a function of urea molarity. Filled markers: our results - blue circles (small system), orange squares (medium system), green triangles (Big system), Medium (SBAM) (light orange pentagones). Unfilled markers: literature data using the same force field - red stars \citenum{weerasinghe2003jpcb}, purple circles \citenum{chibaKirkwoodBuffIntegrals2016}, brown triangles \citenum{milzettiConvergenceKirkwoodBuff2018}, pink squares \citenum{deoliveiraIBITargetingCumulative2016}, yellow crosses \citenum{cortes-huerto2016tjocp}, and blue crossed circles \citenum{cheng2022jcp}. Experimental lines: gray \citenum{chitraMolecularAssociationSolution2002}, light blue \citenum{bowerTHERMODYNAMICSTERNARYSYSTEM1963}. The orange dashed line is a cubic spline interpolation of medium system results.}
\label{fig:ureagammauu}
\end{figure}
Cosolvents in water can significantly alter the thermodynamic properties and stability of proteins through subtle changes in solvation involving proteins, cosolvents, and the solvent. Among cosolvents, urea is particularly well-known for inducing polypeptide unfolding. Consequently, determining thermodynamic and solvation properties in urea-water mixtures is of considerable interest.
The KB force field (KBFF) for the urea-water system~\cite{weerasinghe2003jpcb}, developed based on experimental data~\cite{chitraMolecularAssociationSolution2002}, has been widely used to validate KBI correction methods~\cite{gangulyConvergenceSamplingKirkwood2013,milzettiConvergenceKirkwoodBuff2018,chibaKirkwoodBuffIntegrals2016}, Boltzmann inversion techniques~\cite{deoliveiraIBITargetingCumulative2016}, spatial block analysis methods~\cite{heidari2018mp,cortes-huerto2016tjocp}, and approaches based on structure factor extrapolation~\cite{cheng2022jcp}.

We performed simulations of urea--water mixtures at ten different urea molarities, ranging from 0.5 to 9~\unit{\mol\per\liter}, using the KBFF for urea~\cite{weerasinghe2003jpcb} and SPC/E~\cite{berendsenMissingTermEffective1987} for water. The reference system, labeled ``Medium'', has a volume of approximately \SI{280}{\nano\meter\cubed} and contains around 30,000 atoms. %To assess finite-size effects, some concentrations were also simulated in systems with volumes four times smaller (``Small'', matching the original KBFF study~\cite{weerasinghe2003jpcb}) and twice as large (``Big''). These ``Big'' systems are comparable to those used in more recent studies~\cite{cheng2022jcp,deoliveiraIBITargetingCumulative2016,milzettiConvergenceKirkwoodBuff2018,heidari2018mp}.
To check for possible differences across system sizes, we also simulated selected concentrations in systems with volumes four times smaller (``Small'', matching the original KBFF study~\cite{weerasinghe2003jpcb}) and twice as large (``Big''). The ``Big'' systems are comparable in size to those employed in more recent studies~\cite{cheng2022jcp,deoliveiraIBITargetingCumulative2016,milzettiConvergenceKirkwoodBuff2018,heidari2018mp}.
Density fluctuation spectra were calculated using Eq.~\eqref{eq:fourier_coeff} for each molecular species. The corresponding partial structure factors and direct correlation function analyses are provided in the Supplementary Material, along with fitting results, isothermal compressibility, and partial molecular volumes. We compare our results with available literature data. For method comparison, the ``Medium'' system was also analyzed using the spatial block analysis method as presented in Ref.~\citenum{cortes-huerto2016tjocp}. Results obtained via this method are labeled as Medium (SBAM).

Figure~\ref{fig:gijureawater} presents the resulting KB integrals, alongside literature data obtained using the same force field but different methodologies~\cite{deoliveiraIBITargetingCumulative2016,milzettiConvergenceKirkwoodBuff2018,chibaKirkwoodBuffIntegrals2016}. First, when comparing results using our method from systems of different sizes, no significant discrepancies were observed within the error bars. Second, our values show excellent agreement with previously published data~\cite{deoliveiraIBITargetingCumulative2016,chibaKirkwoodBuffIntegrals2016,milzettiConvergenceKirkwoodBuff2018}. Specifically, for urea concentrations below \SI{4}{\mol\per\liter}, both $G_{uu}$ and $G_{uw}$ closely match the cited studies. At higher molarities, however, our $G_{uu}$ values tend to be higher and $G_{uw}$ lower than those reported in~\cite{milzettiConvergenceKirkwoodBuff2018}, but remains consistent with\citenum{deoliveiraIBITargetingCumulative2016,chibaKirkwoodBuffIntegrals2016}. Compared to experimental values~\citenum{chitraMolecularAssociationSolution2002}, our $G_{uu}$ values are consistently overestimated, while $G_{uw}$ is underestimated at all studied concentrations. This suggests that the KBFF overestimates interactions between similar species, leading to comparatively weaker urea-water interactions.
We further compare the results obtained for the Medium system using our method and the SBAM method results on the same set of data. While the water-water KBIs stay in good agreement, notable discrepancies arise for $G_{uu}$ at low concentrations and for $G_{uw}$ more generally. %Results from SBAM show reasonable agreement with our method but are not fully consistent. Interestingly, the discrepancies between SBAM and the reciprocal-space approach on the same trajectories appear larger than the differences observed between reciprocal-space KBIs computed from trajectories of different system sizes.
Results from SBAM show reasonable agreement with our method but are not fully consistent. Interestingly, the discrepancies between values from SBAM and reciprocal-space approach on the same trajectories appear larger than the differences observed between reciprocal-space values obtained from trajectories of different system sizes.

In Fig.~\ref{fig:ureagammauu}, we report the logarithmic derivative of the urea molar activity $a_u^{\rho} = \rho_{u}\gamma_{u}^{\rho}$, defined as
\[
\gamma_{uu} = \frac{\partial \ln(\rho_{u}\gamma_{u}^{\rho})}{\partial \ln(\rho_u)} = \left(1 + \rho_u (G_{uu} - G_{uw}) \right)^{-1}
\]
Our computed values decrease rapidly to a minimum slightly below 0.85 at $C_u = \SI{4}{\mol\per\liter}$, then increase and cross $\gamma_{uu} = 1$ between \SI{7}{\mol\per\liter} and \SI{8}{\mol\per\liter}. These results are in excellent agreement with previous studies using the same force field~\cite{chibaKirkwoodBuffIntegrals2016,deoliveiraIBITargetingCumulative2016} across the entire concentration range and agree well with~\cite{milzettiConvergenceKirkwoodBuff2018} below $C_u = \SI{4}{\mol\per\liter}$. At higher concentrations, however, our computed derivatives deviate from those reported in~\cite{milzettiConvergenceKirkwoodBuff2018}, and align more closely with~\cite{deoliveiraIBITargetingCumulative2016,chibaKirkwoodBuffIntegrals2016}.

Nonetheless, our values remain systematically lower than the experimental data~\cite{chitraMolecularAssociationSolution2002,bowerTHERMODYNAMICSTERNARYSYSTEM1963}. Given the relationship between $\gamma_{uu}$ and the comparison between computed and experimental KBIs, this discrepancy is expected and consistent with the previously discussed conclusion that the KBFF system exhibits larger composition fluctuations than the experimental system. Values extracted using SBAM on our Medium samples show overall reasonable agreement with our method, although some deviations from the global trend appear unphysical.

Interestingly, although the data across studies were generated under similar conditions, the reported values vary depending on analysis protocols and simulation details. While our computed activity coefficient derivatives differ from both experimental results~\cite{bowerTHERMODYNAMICSTERNARYSYSTEM1963} and from those reported by other simulation studies~\cite{heidari2018mp,cortes-huerto2016tjocp,cheng2022jcp}, they are consistent with results from alternative approaches~\cite{chibaKirkwoodBuffIntegrals2016,deoliveiraIBITargetingCumulative2016,milzettiConvergenceKirkwoodBuff2018}.

The origin of these discrepancies remains unclear, especially given that all studies use the same force field. Potential contributing factors may include differences in equilibration protocols, electrostatic calculation method, choices in cutoff distances, applied energy and pressure corrections, or even the choice of thermostat and barostat. Each of these elements can subtly affect long-range correlations and potentially influence Kirkwood-Buff integral estimates and derived thermodynamic quantities.

\subsection{sodium chloride aqueous mixture}
\begin{figure}
  \includegraphics[width=\columnwidth]{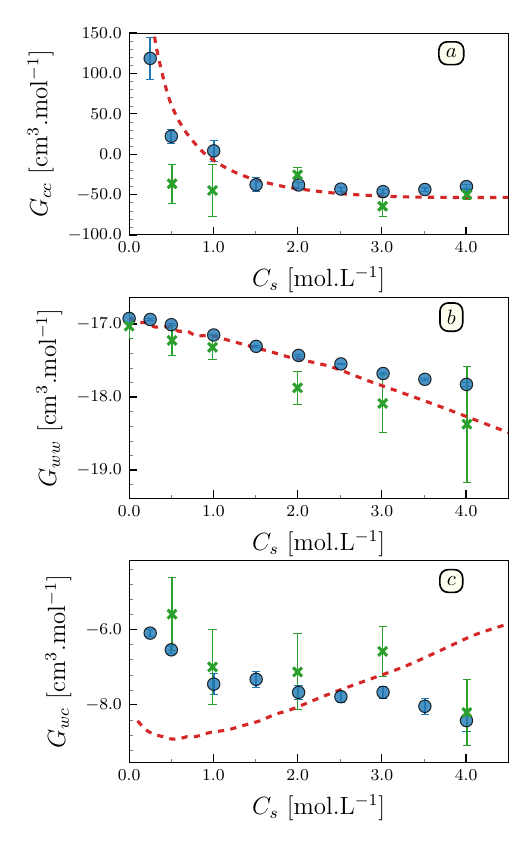}
  \caption{Kirkwood–Buff integrals as a function of NaCl molality. (a) Cosolvent–Cosolvent, (b) Water–Water, (c) Cosolvent–Water. Stars: \citealt{weerasinghe2003jcp}. Red dashed line: experimental values from \citealt{chitraMolecularAssociationSolution2002}.}
  \label{fig:gijnacl}
\end{figure}

\begin{figure}
  \includegraphics[width=\columnwidth]{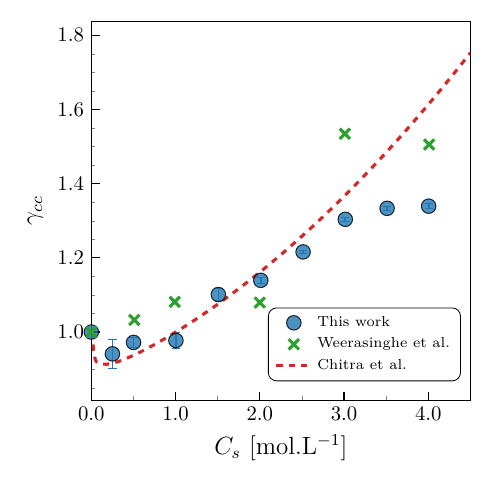}
  \caption{Molar activity coefficient derivative of NaCl as a function of molarity. Blue circles: this work. Crosses: \citealt{weerasinghe2003jcp}. Red line: \citealt{chitraMolecularAssociationSolution2002}.}
  \label{fig:naclgammacc}
\end{figure}

In salt solutions, the electroneutrality condition imposes a constraint on the number of anions and cations such that \( \sum_i N_i q_i = 0 \) where $i$ stand for ionic species, $q_i$ is the charge of the ions $i$ and $N_i$ the number of ions $i$. Due to this constraint, the mean numbers of anions and cations are coupled in the grand canonical ensemble, which restricts their cross-fluctuations. As a result, the symmetric structure factor matrix \( S \) becomes singular, with zero determinant, and is therefore non-invertible~\cite{friedman1970jpc}. In other words, the constraint on ion numbers in the grand canonical ensemble leads to a dependency between the chemical potentials of charged species in the canonical ensemble, preventing a well-defined Legendre transform between the two~\cite{kusalik1987tjocp,ben-naim2006,oconnell1975jsc,behera1998jcp}.

To overcome this issue, the \( S \) matrix, originally defined on a species basis (solvent, cation, anion), can be transformed to a neutral salt or cosolvent basis (solvent and total ion number), reducing the matrix rank and eliminating the singularity~\cite{oconnell1975jsc}. This enables a direct link between microscopic density fluctuations of cosolvent and thermodynamic properties~\cite{oconnell1975jsc,kusalik1987tjocp} and allows the extraction of macroscopic properties without referencing to individual ionic species behavior~\cite{oconnell1999fpe,behera1998jcp}. In cosolvent basis, due to mean electroneutrality of cosolvent in large volume, the low-q part of direct correlation function are determined between two overall neutral species, leading to a regular polynomial expansion~\cite{rovere1979pmb,giaquintaHydrodynamicCorrelationFunctions1976a,kjellander2005jpamg}.
We therefore treat the pseudo-ternary Water/$\text{Na}^+$/$\text{Cl}^-$ solution as a binary mixture of water (w) and indistinguishable ions or cosolvents (c). For a salt of stoichiometry \( n_+ : n_- \) and total number of ions \( n = n_+ + n_- \), the cosolvent concentration \( \rho_c \) and molarity \( C_c \) relate to the classical salt values as follows: \( C_c = n C_s \), \( \rho_c = n \rho_s \), \( V_s = n V_c \), and \( \gamma_c = \gamma_{\pm} \). Additionally, one has: 
\[
d\mu_s = n\,d\mu_c, \quad \rho_s\,d\mu_s = \rho_c\,d\mu_c, \quad d\ln(\rho_s) = d\ln(\rho_c),
\]
\[
\rho_s V_s + \rho_w V_w = \rho_c V_c + \rho_w V_w = 1,
\]
\[
\rho_c\,d\ln(a_c) + \rho_w\,d\ln(a_w) = \rho_s\,d\ln(a_s) + \rho_w\,d\ln(a_w) = 0
\]
at constant pressure and temperature.

We used the KBFF force field~\cite{weerasinghe2003jcp} for NaCl and the SPC/E model for water. Systems were simulated for salt molarities \( C_s =0.25, 0.5, 1, 1.5, 2, 2.5, 3, 3.5, 4\;\si{\mol\per\liter} \) using cubic boxes of approximately \SI{6.5}{\nano\meter} side length, containing about 9500 water molecules. The number of cosolvent ions varied from 84 to 1556 depending on the target molarity. We compute the density fluctuation in reciprocal space for cosolvents and water and determine the partial structure factor between cosolvents, water and cross-correlation between cosolvents and water. Obtained partial structure factor and fourier components of direct correlation functions are presented in the Supplementary Material with corresponding fits. The derived thermodynamic quantities are compared with the original KBFF results~\cite{weerasinghe2003jcp} and with experimental data interpreted using the Kirkwood–Buff formalism~\cite{chitraMolecularAssociationSolution2002}, from which the KBFF parameters were originally fitted.

As shown in Fig.~\ref{fig:gijnacl}-a, our KBIs between cosolvent particles agree significantly better with experimental data at low concentrations than the original KBFF values. At higher molarities, our results remain consistent with both KBFF and experimental references. The cosolvent–water KBIs (Fig.~\ref{fig:gijnacl}-c) also follow the trend of the original KBFF, and their magnitudes remain comparable to the experimental data. However, unlike the experimental data which exhibit a minimum around \SI{0.5}{\mol\per\liter} followed by an increase, our KBIs show a monotonic decrease with increasing molarity. The water–water KBIs (Fig.~\ref{fig:gijnacl}-b) closely match both the original and experimental results up to \SI{3}{\mol\per\liter}, beyond which our values plateau.

The derivative of the cosolvent molar activity \( \gamma_{cc} = \left[1 + \rho_c (G_{cc} - G_{cw}) \right]^{-1} \), shown in Fig.~\ref{fig:naclgammacc}, is also in closer agreement with experimental values than the original KBFF data. This improvement stems from our \( G_{cc} \) values aligning more closely with experimental results. At concentrations below \SI{3}{\mol\per\liter}, the agreement is satisfactory. At higher molarities, deviations are observed, likely due to the values of \( G_{wc} \), which causes a corresponding plateau in $\gamma_{cc}$.

%In conclusion, by using the cosolvent framework to calculate density fluctuation in Fourier space, it remains possible to determine the Fourier components of direct correlation function and using the same asymptotic expansion as the one applied to LJ system, extract reliable thermodynamic parameters via KB inversion. We emphasize that accurate results can be obtained even with relatively small numbers of ions and moderate system sizes using the standard cosolvent formalism without applying any finite-size corrections. This latter point may be of particular interest in the field of salts solvation. Recently, combining KB theory and free energy calculations\cite{chattopadhyayDeterminationAqueousSolubility2025} permit to determine the salt solubility limit of various aqueous sodium chloride model. However, as stated by the authors chemical potential of solutions are highly sensitive to mandatories finite-size effects corrections, the reciprocal-space approach may offer a valuable complementary or alternative methodology due to its inherently lower sensitivity to finite-size effects.

In conclusion, by employing the cosolvent framework to evaluate density fluctuations in Fourier space, it remains possible to determine the Fourier components of direct correlation function and using the same asymptotic expansion as the one applied to LJ system, extract reliable thermodynamic parameters via KB inversion. Notably, accurate results can be obtained even for relatively small numbers of ions and moderate system sizes, using the standard cosolvent formalism without applying explicit finite-size corrections. This feature may be particularly valuable in the context of salt solvation studies. For example, a recent study by~\citealt{chattopadhyayDeterminationAqueousSolubility2025} combined Kirkwood–Buff theory with free energy calculations to determine the solubility limits of various aqueous sodium chloride models. However, the authors emphasized that accurate determination of chemical potentials using KB route is highly sensitive to the inclusion of different finite-size corrections. In this context, our reciprocal-space approach may offer a useful complementary or alternative methodology, owing to its intrinsically lower sensitivity to finite-size artifacts.

\section{Conclusion}
We develop and validate an extension of the methodology proposed by \citealt{nichols2009pre}, which addresses the inversion of the KB relationships and enables access to derivatives of thermodynamic properties using standard partial structure factor, calculated in canonical ensemble, supported with linear algebra formulation of KB inversion. Through a series of systems with varying physical interactions, from binary Lennard-Jones mixtures to electrolyte solutions, we aim to demonstrate that, based solely on the wavelength dependence of the Fourier components of the direct correlation functions between components, it is possible to extract macroscopic thermodynamic derivatives via a simple extrapolation using established relationships without applying finite size correction.
Because the direct correlation functions in real space decay more rapidly than the pair correlation function (or indirect pair correlation functions), their Fourier components exhibit a more monotonic behavior at low-$q$. This improved regularity makes low-$q$ extrapolation easier, more reliable and ultimately leads to more accurate thermodynamic estimates.% Those advantages enable access to thermodynamic limits in systems up to ten times smaller than those typically required by spatial block analysis methods or using corrected radial distribution function integrals. 
%These advantages allow accurate estimation of thermodynamic quantities from systems that are significantly smaller, by up to an order of magnitude, than those typically required by real-space methods such as SBAM or RDF integration, particularly in compositionally asymmetric regimes.
These advantages make it possible to obtain reliable estimates of thermodynamic quantities from systems that are substantially smaller than those typically required by real-space methods such as SBAM or RDF integration. In our tests, reciprocal-space analysis provided consistent results across systems differing by up to an order of magnitude in size, particularly in compositionally asymmetric regimes.
The case of the hexane–ethanol mixture provides a compelling illustration of the effectiveness of the proposed method, particularly in systems with large density fluctuations, where convergence in real space demands larger system sizes.

Although assumptions must be done regarding the low-$q$ expansion of the Fourier transform of the direct correlation function, no additional corrections are required. In practice, a simple power series expansion in $q$, using even powers and including a $q^3$ term accounting for van der Waals interaction, appears sufficient to accurately capture the low-$q$ behavior of the direct correlation function for all the studied systems. This approach yields precise macroscopic results across a wide range of systems, including Lennard-Jones mixtures, polar molecular mixtures, and even charged aqueous solutions far from simple site-site van der Waals interaction.

To conclude, we note the recent introduction of the ``countoscope'' technique~\cite{mackayCountoscopeMeasuringSelf2024,hoangngocminh2023fd}. Briefly, this method evaluates density (or charge) fluctuations in Fourier space, thus applying a weighting to the Fourier components permits to mimics the effect of reducing the box size. An inverse Fourier transform of the weighted components then permits to reach correlations for systems of a specified size. By analyzing the system size evolution of the correlations, one can probe the hyperuniformity length and extrapolate beyond this length to macroscopic values.
This technique might offer two major advantages: (1) the Fourier components of the subvolumes are computed using the complete information from the full simulation box, resulting in improved statistical accuracy and the elimination of finite-size effects of volume integration; (2) no assumptions are needed regarding the low-$q$ asymptotic behavior of the direct correlation function. Instead of relying on a fitting procedure, this method performs an extrapolation to the infinite system size using the framework of small-system thermodynamics.

\section*{Supplementary Material}
The supplementary material contains discussions on the Equation of state formulation used for binary and quaternary Lennard-Jones system.
It also contains data issued from KB inversion methodology not discussed in the article.

\section*{Data Availability Statement}
The data that support the findings of this study are available within the article and its supplementary material.
The data that support the findings of this study are openly available in Zenodo at http://doi.org/10.5281/zenodo.15591522

%\nocite{*}
\bibliography{Biblio_KBI}

\end{document}

\title{Supplementary Material - Robust Kirkwood-Buff Inversion in Complex Mixtures via Reciprocal-Space Methods}
\author{R.Busselez}
 \email{remi.busselez@univ-lemans.fr}
 \affiliation{Institut des Molécules et Matériaux du Mans, UMR CNRS 6283,
Le Mans université, 72085 Le Mans, France} 

\date{\today}% It is always \today, today,
\maketitle

\section{Equation of State for Lennard Jones mixture}
To determine the theoretical properties of a mixture of Lennard-Jones particles, we employ the modified Benedict–Webb–Rubin equation of state (mBWR EoS) tailored for the Lennard-Jones potential. The 33 parameters used in this EoS were obtained from Refs.~\cite{mayRiemannianGeometryStudy2012, mayErratumRiemannianGeometry2012}. This formulation accurately captures the thermodynamic behavior of the 12-6 Lennard-Jones (LJ) fluid and reliably represents the essential physical characteristics of simple, pure fluids.

For mixtures, a conformal solution theory or a corresponding states approach is required~\cite{hansenTheorySimpleLiquids2013}. Among these, the van der Waals one-fluid model (vdW1) is recognized as the most effective approximation for LJ mixtures~\cite{lelandStatisticalThermodynamicsMixtures1968} assuming that size and energy interaction disparities between species are not too large. Within this framework, the effective parameters for the mixture are computed using the following mixing rules:
\begin{equation}
  \begin{aligned}
    \sigma_x^3 &= \sum_i \sum_j x_i x_j \sigma_{ij}^3 \\
    \epsilon_x \sigma_x^3 &= \sum_i \sum_j x_i x_j \epsilon_{ij} \sigma_{ij}^3
  \end{aligned}
\end{equation}
Here, $\sigma_{ij}$ and $\epsilon_{ij}$ denote the size and energy parameters, respectively, of the 12-6 LJ potential between species $i$ and $j$.

To compute the thermodynamic properties of a mixture from the mBWR EoS at a given pressure $P$ and temperature $T$, the following procedure is applied. For each desired composition, we first calculate the effective one-fluid parameters $\sigma_x$ and $\epsilon_x$. These are then used to define the reduced pressure and temperature via the following rescaling relations:
\begin{equation}
  P^* = P \frac{\sigma_x^3}{\epsilon_x}, \quad T^* = \frac{k_B T}{\epsilon_x}
\end{equation}

Using these reduced variables, the reduced density $\rho^*$ is obtained by minimizing the mBWR EoS pressure function $P^*(\rho^*, T^*)$. From the reduced density, temperature, and pressure, we compute various reduced thermodynamic properties such as the isothermal compressibility, residual Gibbs free energy, entropy, and enthalpy~\cite{johnsonLennardJonesEquationState1993}, which are then converted to real physical units using $\sigma_x$ and $\epsilon_x$.

Partial molar volumes, chemical potentials of the individual species, and the thermodynamic factor matrix are evaluated using unconstrained numerical derivatives (Jacobian and Hessian matrices) of the residual Gibbs free energy with respect to molar composition~\cite{taylor1991cec}. Finally, the Kirkwood–Buff integrals are calculated based on the obtained values of isothermal compressibility, partial molar volumes, and the thermodynamic factor.

\section{Binary Lennard Jones Mixture}
We report the isothermal compressibility, excess volume, and partial molar volumes obtained from KB analysis of binary Lennard-Jones mixtures in NVT simulations. The isothermal compressibility is also compared with values obtained from volume fluctuations in NPT simulations, using the relation:

\[
\kappa_T = \frac{\langle (V - \langle V \rangle)^2 \rangle}{k_B T \langle V \rangle}
\]

\begin{figure}
    \centering
    \includegraphics[width=\columnwidth]{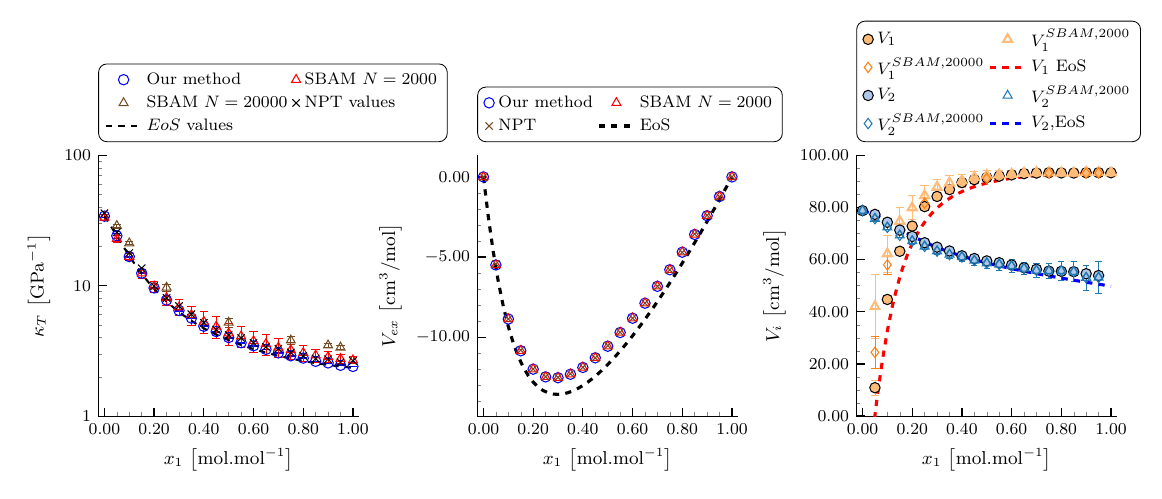}
    \caption{
    Left: Isothermal compressibility (log scale) as a function of the molar fraction of particle 1. Blue circles: reciprocal-space analysis ($N=2000$); black crosses: NPT volume fluctuation;Red and Brown triangles: SBAM analysis on respectively $N=2000$ and $N=20000$ ; dashed line: mBWR/vdW1 EoS.\\
    Center: Excess volume vs. molar fraction. Blue circles: Reciprocal-space analysis ($N=2000$); brown crosses: values from NPT simulation; Red triangle: SBAM analysis ($N=2000$) ;dashed line: mBWR/vdW1 EoS.\\
    Right: Partial molar volume vs. molar fraction. Orange circle: Species 1, Reciprocal space analysis ($N=2000$); Orange triangle: Species 1 SBAM Analysis ($N=2000$); Orange diamond: Species 1, SBAM analysis ($N=2000$); Red dashed line: Species 1 from mBWR EoS; blue circles: Species 2, Reciprocal space analysis ($N=2000$); blue triangle: Species 2, SBAM analysis ($N=2000)$; blue diamond: Species 2, SBAM analysis ($N=2000$); Blue dashed line: Species 2 mBWR EoS 
    /dashed line: Particle 2 from KB/mBWR.
    }
    \label{fig:si-binarythermo}
\end{figure}

Additionally, we compare the results from simulations obtained with our proposed reciprocal-space analysis and Spatial Block Averaging Method with predictions from the mBWR/vdW1 EoS. These comparisons are summarized in Figure~\ref{fig:si-binarythermo}.

The isothermal compressibility values obtained from KB inversion agree well with both the mBWR/vdW1 EoS and NPT fluctuation analysis. Although all methods of KB inversion gives satisfactorily agreement, we must emphasize that our proposed method gives values closer to the EoS. 

Partial molar volumes obtained from KB inversion in the NVT ensemble are in overall good agreement with mBWR/vdW1 predictions for mole fractions \(x_1\) between 0.4 and 0.7. At low (respectively high) values of \(x_1\), the partial molar volume of species 1 (respectively species 2) is slightly higher in the KB results than in the mBWR/vdW1 EoS.
Values of partial volume obtained using our proposed reciprocal space analysis for KB inversion are in all cases closer to EoS predictions than SBAM analysis. Using SBAM analysis, the increase in system size tends to align more closely the results to EoS but results remains significantly differents for $V_1$ at low $x_1$ fraction. As reported in the article, our proposed method is in closer agreement to EoS at any molar fraction.

Excess volumes in NPT simulations are computed from the average volume at each composition, using the standard expression:

\[
V_{\text{ex}}(x_1) = V(x_1) - \left[ x_1 V(x_1=1) + (1 - x_1) V(x_1=0) \right]
\]

We also compute excess volume from the partial molar volumes obtained via KB inversion in the NVT ensemble:

\[
V_{\text{ex}}(x_1) = x_1 \left[ V_1(x_1) - V_1(x_1=1) \right] + (1 - x_1) \left[ V_2(x_1) - V_2(x_1=0) \right]
\]

The excess volumes obtained from both methods are superimposed across the full composition range. However, both slightly deviate from the values predicted by the mBWR/vdW1 model. We thus conclude that the van der Waals one-fluid approximation is in close agreement with our simulation data and provides a reliable reference model for comparison of the  proposed method.

\section{Quaternary Lennard Jones Mixture}
\begin{figure}
\includegraphics[width=0.75\columnwidth]{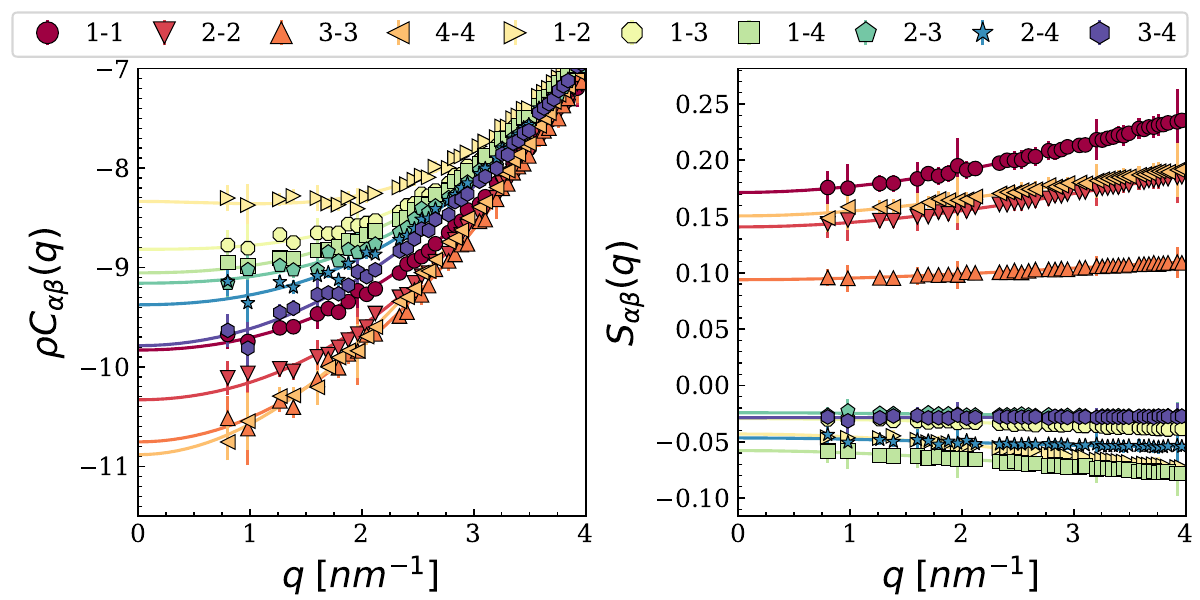}
\caption{
  Left: Fourier components of direct correlation functions between species of quaternary lennard jones mixture. Species are given in legend. The lines are the fit. Right: Partial structure factor obtained from simulations. Lines are the $S_{\alpha-\beta}(q)$ values obtained from the polynomial fit of $C_{\alpha-\beta}(q)$\label{fig:si-cqsqquat}}
\end{figure}

\begin{figure}
  \includegraphics[width=0.95\textwidth]{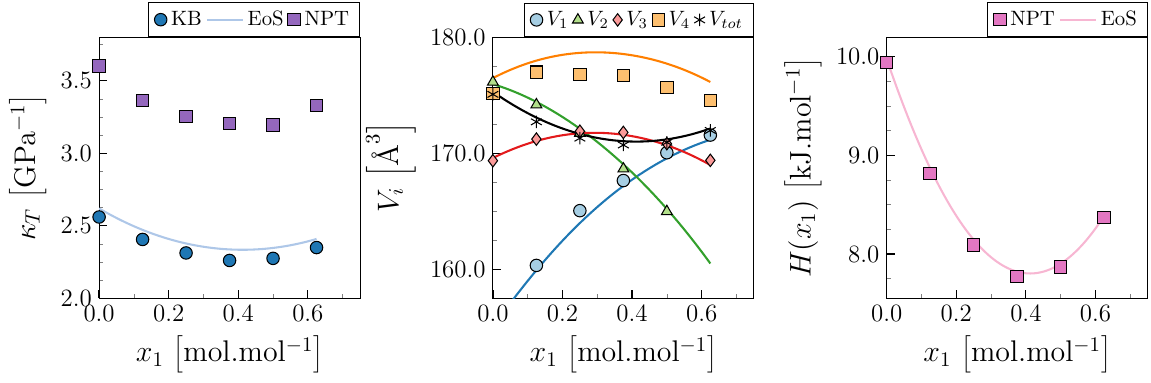}
  \caption{Left: Isothermal compressibility of the quaternary mixture as a function of species 1 molar fraction.  Circle are data from KB analysis, Line is mBWR/vdW1 EoS, Squares is the volume fluctuation of NPT simulations.
  Center: Partial volumes of the species as a function of species 1 molar fraction. Filled Markers KBI analysis; Line mBWR/vdW1 EoS; asterisk is the mean atomic volume and black line the mean atomic volume from EoS.
  Right: Enthalpy of quaternary mixture as a function of species 1 molar fraction. Square: NPT simulation; line: mBWR/vdW1 EoS}
  \label{fig:si-kvquat}
\end{figure}

In Fig.~\ref{fig:si-cqsqquat}, we present, as a representative example, the partial structure factors between species for the quaternary mixture with composition \(x_1 = 0.375\), \(x_2 = 0.25\), \(x_3 = 0.125\), and \(x_4 = 0.25\). We also show the Fourier components of the direct correlation functions \(C_{ij}(q)\) along with their corresponding fits based on the asymptotic low-\(q\) approximation. The calculated partial structure factor \(S(q)\), obtained from the asymptotic expansion of \(C(q)\), is superimposed on the partial structure factor data to demonstrate consistency.

Figure~\ref{fig:si-kvquat} (left panel) displays the isothermal compressibility extracted from Kirkwood-Buff (KB) analysis and volume fluctuations in NPT simulations, alongside the compressibility predicted by the mBWR/vdW1 EoS. The KB-derived compressibility values closely follow the EoS trend, showing very good agreement overall, with a systematic offset of approximately \SI{0.05}{\per\giga\pascal} towards lower values. In contrast, compressibility values derived from volume fluctuations in NPT simulations are significantly overestimated, by about \SI{1}{\per\giga\pascal}. This overestimation may be due to the proximity of the system to the supercritical domain inducing huge fluctuations in volume box and subsequent difficulties to determine isothermal compressibility.
The center panel of Fig.~\ref{fig:si-kvquat} reports the partial molar volumes of each species, compared with those obtained from the mBWR/vdW1 EoS. Good agreement is observed for species 1 through 3. For species 4, the KB-derived partial volume is systematically lower than the EoS prediction. Nonetheless, our KB analysis remains fully consistent with the EoS results within the framework of the one-fluid model approximation.
Finally the right panel reports the total enthalpy (residual and ideal gaz) of the system as a function of $x_1$ species for the NPT simulation and EoS of the quaternary mixture. Values are in perfect agreement underlining the reliability of the Van Der Waals one fluid model to determine thermodynamic properties of the quaternary mixture of LJ particles.

\section{Hexane-Ethanol}
\begin{figure}
\includegraphics[width=\columnwidth]{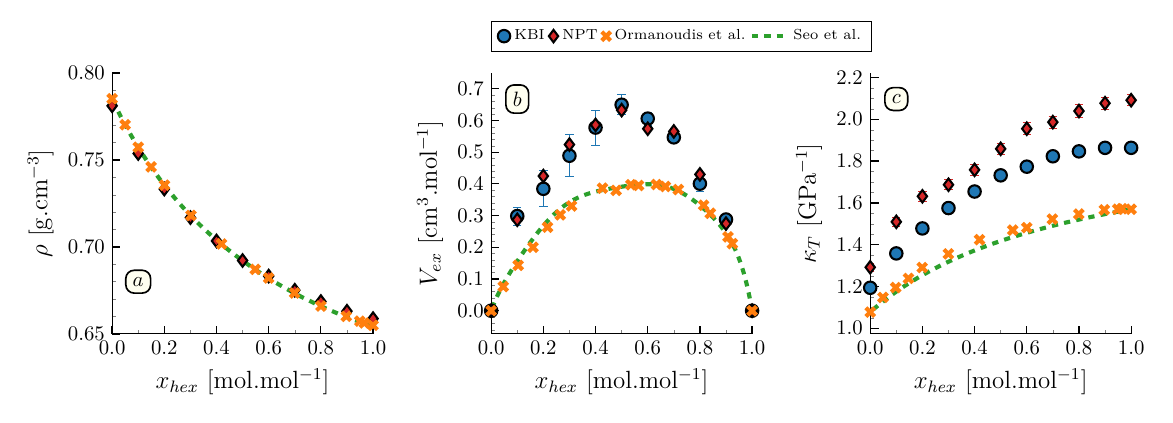}
\caption{Blue Circles: Data extracted from KB inversion method on NVT system. Red Diamonds: Data from NPT simulation. Orange crosses: Experimental data from \citealt{ormanoudisVolumetricPropertiesBinary1991}. Green dashed line: Fitted experimental values of \citealt{seo2024jpcb}.
Figure a: Density of ethanol-hexane mixture as a function of hexane molar fraction. Figure b: Excess Volume of mixture as a function of hexane molar fraction. Figure c: Isothermal compressibility of mixture as a function of hexane molar fraction.}
\label{fig:si-vhexeth}
\end{figure}

\begin{figure}
  \includegraphics[width=\columnwidth]{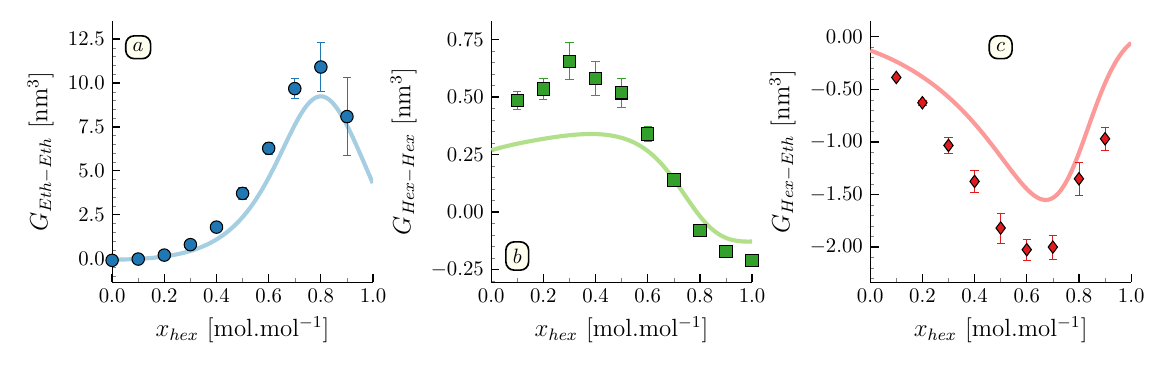}
  \caption{KB integrals obtained for Hexane-Ethanol mixture as a function of hexane molar fraction. Marks: Our data obtained using KB inversion. Lines: Estimation obtained using experimental data, following the method of \citealt{seo2024jpcb} and a Wilson model of mixture. Figure a: KB Integral between ethanol molecules, Figure b: KB integral between hexane molecules, Figure c: KB integral between hexane and ethanol molecules.}
  \label{fig:si-gijhexeth}
\end{figure}

We provide additional results for the hexane–ethanol system in Fig.~\ref{fig:si-vhexeth}. Values estimated from our KB inversion analysis are compared with the experimental data reported by \citealt{ormanoudisVolumetricPropertiesBinary1991}, including density, isothermal compressibility, and excess molar volume, all measured at the same temperature and pressure conditions: \(T = \SI{298.15}{\kelvin}\) and \(P = \SI{1}{\bar}\).

At first glance, the density behavior of the mixture closely matches the experimental data~\cite{ormanoudisVolumetricPropertiesBinary1991}. Specifically, our simulations yield a slightly higher density for pure ethanol and a slightly lower density for pure hexane, in agreement with previous findings~\cite{petris2019jpcb}. However, these small differences result in an excess molar volume that is significantly overestimated in our analysis, by as much as 50\% relative deviation near \(x_{\text{hex}} = 0.5\), indicating a greater deviation from ideal mixing behavior than that observed experimentally.

The isothermal compressibility values obtained from volume fluctuations in NPT simulations are consistently higher than those derived via KB inversion in the NVT ensemble across the entire composition range. This discrepancy becomes more pronounced at higher hexane mole fractions.

Both simulation-derived compressibility curves qualitatively agree with experimental trends: the overall shape of the isothermal compressibility as a function of hexane concentration is captured. However, the absolute values deviate, with discrepancies between 10\% and 20\% depending on the hexane mole fraction, when compared to compressibilities obtained via KB inversion.

In Fig.~\ref{fig:si-gijhexeth}, the KB integrals obtained from our inversion method are compared with those estimated following the procedure described by \citealt{seo2024jpcb}, which is based on experimental data. In that approach, KB integrals are calculated from fitted experimental measurements, requiring input for isothermal compressibility, density, excess volume, and the thermodynamic factor, with mixing laws applied throughout. The thermodynamic factor is determined using experimental data from \citealt{hongo1994jced} and a Wilson model, which shows near-perfect agreement with our simulation results, as discussed in the main text.

However, due to the observed differences in excess volume and isothermal compressibility between simulation and experimental data, the final agreement for the KB integrals remains only semi-quantitative.

\section{Urea Water mixture}

% \pgfplotstabletypeset[
%   col sep=semicolon,
%   %%col sep=&,
%   %%row sep=\\,
%   every head row/.style={
%     before row={
%       \hline
%        & \multicolumn{3}{|c}{Small} &\multicolumn{3}{|c|}{Medium} &\multicolumn{3}{c|}{Big}\\
%     },
%     after row=\hline,
%   },
%   every last row/.style={
%     after row=\hline
%   },
%   column type/.add={|}{|},
%   string replace*={nan}{},
%   set thousands separator={},
%   %empty cells with={X},
%   precision=3,sci zerofill,
%   columns/Cs/.style={column name=Cs (target)},
%   columns/NwS/.style={column name=Water},
%   columns/NuS/.style={column name=Urea},
%   columns/Csrs/.style={column name=Cs obtained},
%   columns/NwM/.style={column name=Water},
%   columns/NuM/.style={column name=Urea},
%   columns/Csrm/.style={column name=Cs obtained},
%   columns/NwB/.style={column name=Water},
%   columns/NuB/.style={column name=Urea},
%   columns/Csrb/.style={column name=Cs obtained},
% ]{./Figures/UreaWater/System_Prop.csv}

\begin{figure}
  \includegraphics[width=\columnwidth]{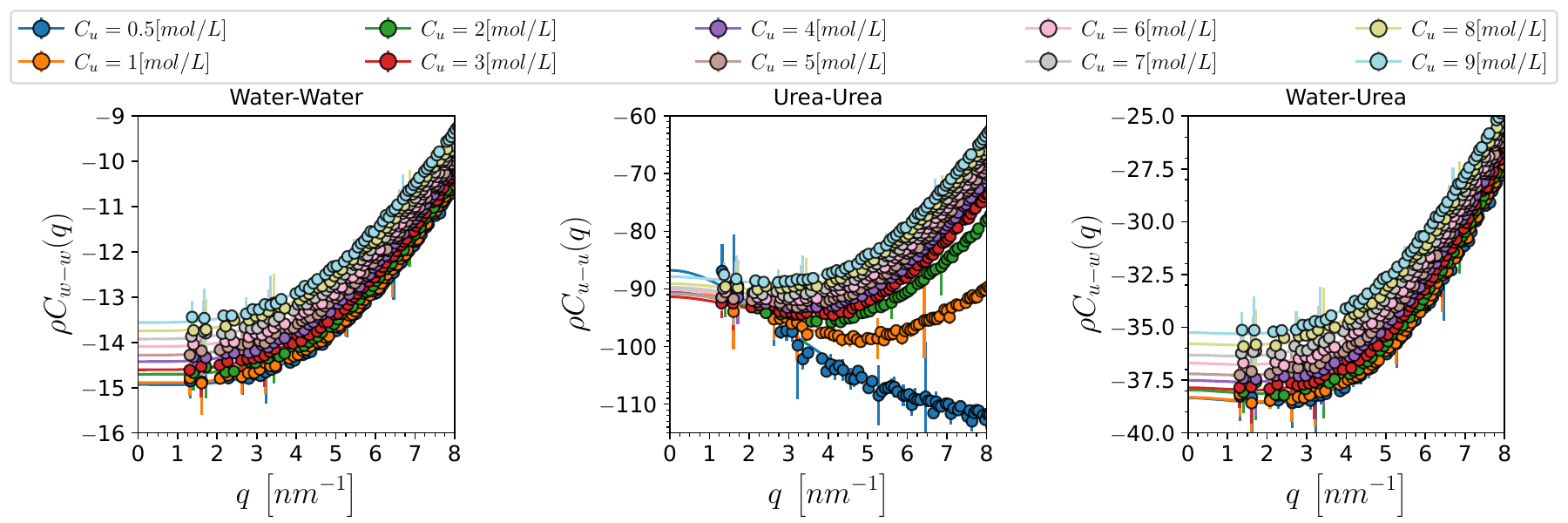}
  \caption{Fourier components of direct correlation functions as a function of wavevector for different urea molarity.
  Marks are the calculated values, lines are fit using asymptotic law.
  Left: Water-Water
  Middle: Urea-Urea
  Right: Water-Urea
  }
  \label{fig:si-cqurea}
\end{figure}

\begin{figure}
  \includegraphics[width=\columnwidth]{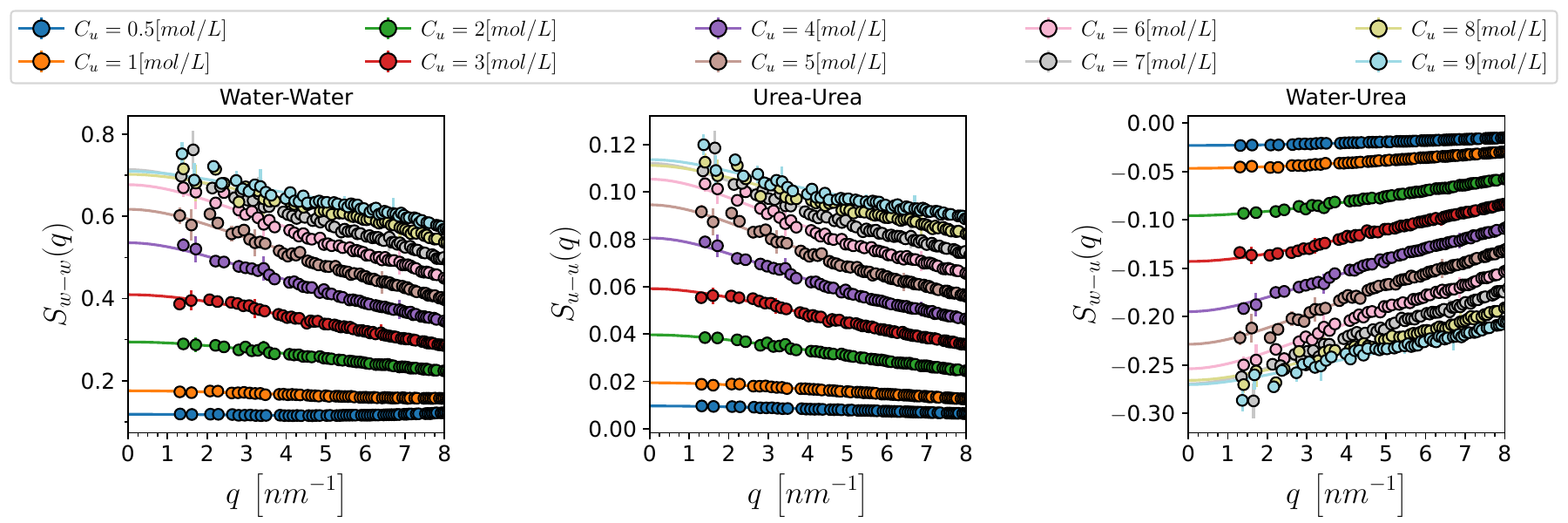}
  \caption{Partial structure factor as a function of wavevector for different urea molarity.
  Marks are the calculated values, lines provide from the fit of fourier transform of direct correlation functions.
  Left: Water-Water
  Middle: Urea-Urea
  Right: Water-Urea
  }
  \label{fig:si-squrea}
\end{figure}
In Fig.~\ref{fig:si-cqurea}, we report the values of \(\rho C(q)\) for the various urea molarities studied. The corresponding fits and extrapolations to the low-\(q\) region are superimposed. Figure~\ref{fig:si-squrea} presents the partial structure factors calculated directly from simulations, alongside the partial structure factor reconstructed from the fitted direct correlation functions in reciprocal space. We observe that the asymptotic form used for extrapolation is fully consistent with the behavior of the data across all urea concentrations.

\begin{figure}
\includegraphics[width=\columnwidth]{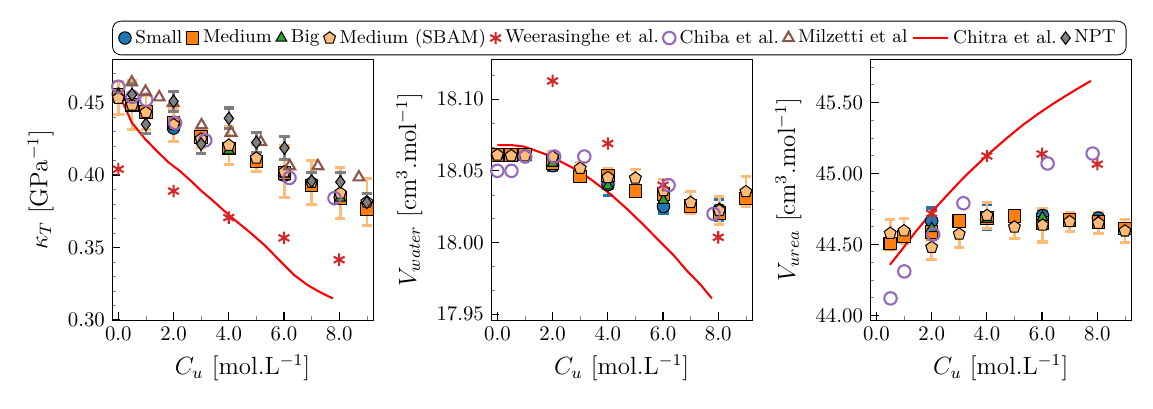}
\caption{Left: Isothermal compressibility as a function of urea molarity.
Middle: Partial molar volume of water as a function of urea molarity. 
Right: Partial molar volume of urea as a function of urea molarity. 
In all panels, filled markers represent our results: blue circles correspond to the small system, orange squares to the medium system, green triangles to the large system, light orange pentagons to the medium system analyzed using SBAM, and gray diamonds to values obtained from volume fluctuations in NPT simulations. Empty symbols represent other simulation data under the same conditions. Red stars: \citealt{weerasinghe2003jpcb}, purple circles: \citealt{chibaKirkwoodBuffIntegrals2016}, brown triangles: \citealt{milzettiConvergenceKirkwoodBuff2018}. The red line represents experimental data from \citealt{chitraMolecularAssociationSolution2002}.}
\label{fig:si-kappavmurea}
\end{figure}

Figure~\ref{fig:si-kappavmurea} summarizes the isothermal compressibility and the partial molar volumes of both water and urea, as obtained through KB inversion using our method for three different system sizes. We also include our values derived via the spatial block analysis method (SBAM) applied to the medium-sized system. These results are compared against previously published simulation data using the same force field, as well as available experimental measurements.

Importantly, the KB inversion results appear independent of system size, within the limits of the method's uncertainty. Since our small system is comparable in size to that used in the foundational study by \citealt{weerasinghe2003jpcb}, we conclude that reliable values can be extracted even from moderately sized simulations. Additionally, our SBAM analysis yields consistent results, further validating the robustness of our approach.

Our values for isothermal compressibility and the partial molar volume of water are in excellent agreement with those reported by \citealt{chibaKirkwoodBuffIntegrals2016} and \citealt{milzettiConvergenceKirkwoodBuff2018}, but they deviate notably from the data of \citealt{weerasinghe2003jpcb}. This trend mirrors the discrepancies observed for the KB integrals and the derivatives of molar activity coefficients. Furthermore, our results show systematic differences compared to the experimental values reported by \citealt{chitraMolecularAssociationSolution2002}.

Interestingly, the partial molar volume of urea obtained in our analysis remains relatively constant across the range of concentrations studied, in contrast to the more pronounced variations reported in other works.

\section{NaCl aqueous mixture}

\begin{figure}
\includegraphics[width=\columnwidth]{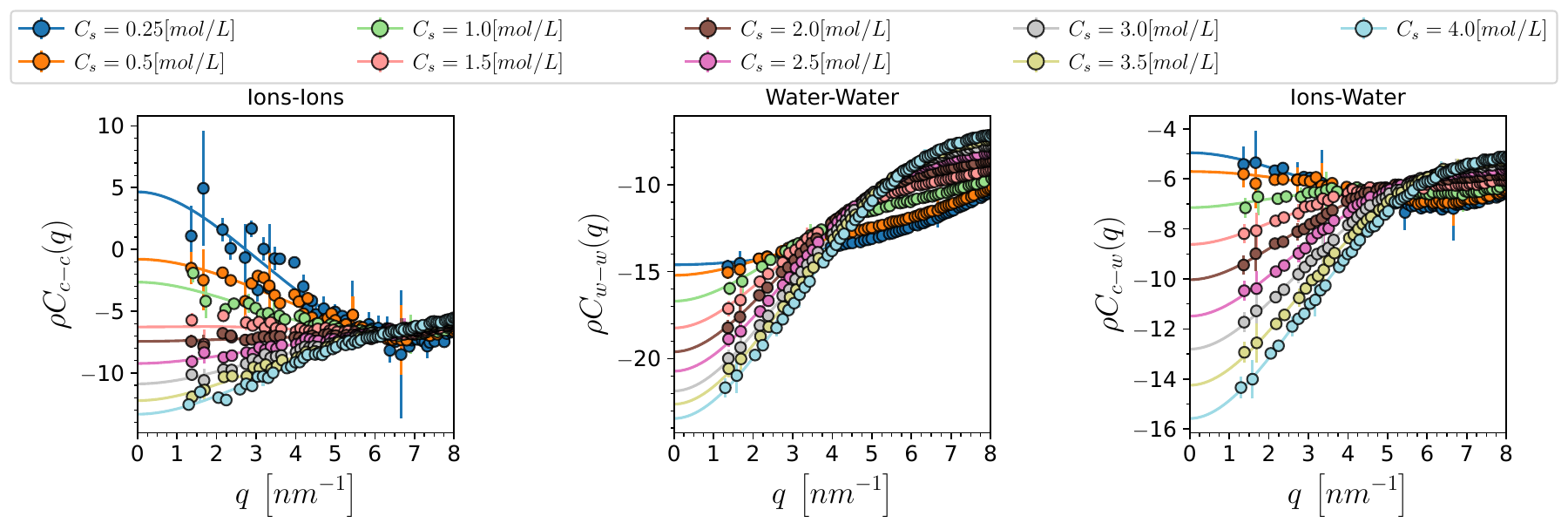}
\caption{Fourier components of direct correlation functions as a function of wavevector for different salt molarity.
  Marks are the calculated values, lines are fit using asymptotic law.
  Left: Cosolvents-Cosolvents
  Middle: Water-Water
  Right: Water-Cosolvents
  }
  \label{fig:si-cqnacl}
\end{figure}
\begin{figure}
\includegraphics[width=\columnwidth]{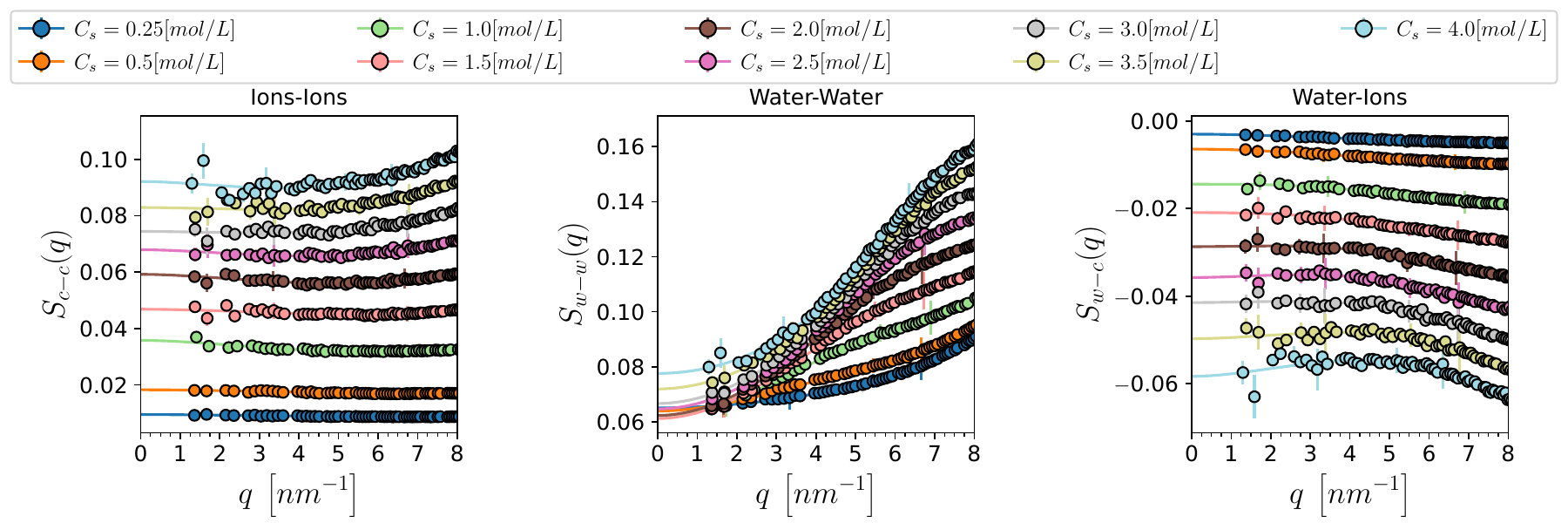}
  \caption{Partial structure factor as a function of wavevector for different salt molarity.
  Marks are the calculated values, lines provide from the fit of fourier transform of direct correlation functions.
  Left: Cosolvents-Cosolvents
  Middle: Water-Water
  Right: Water-Cosolvents
  }
  \label{fig:si-sqnacl}
\end{figure}

\begin{figure}
  \includegraphics[width=\columnwidth]{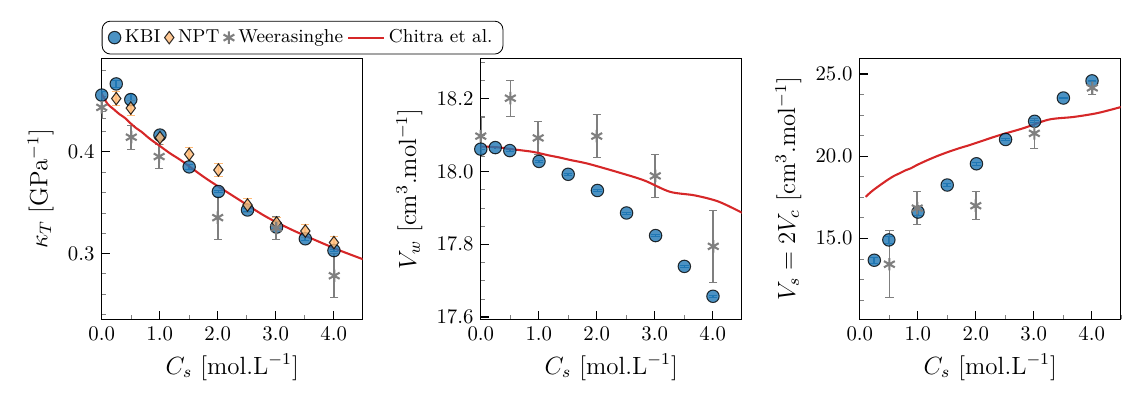}
  \caption{Left: Isothermal compressibility as a function of NaCl salt molarity. Middle: Water partial molar volume as a function of NaCl salt molarity. Right: Urea partial molar volume as a function of NaCl salt molarity.
Blue circles: KBI analysis, this work. Orange Diamond:Fluctuations NPT. Gray Stars:\citealt{weerasinghe2003jcp}, Red line are experimental values from \citealt{chitraMolecularAssociationSolution2002}.}
\label{fig:si-kvmnacl}
\end{figure}
We represent in fig.\ref{fig:si-cqnacl} the $\rho C(q)$ values for the different molarity of NaCl aqueous solution studied. We superimpose the fit and extrapolation obtained at low q values. In fig.\ref{fig:si-sqnacl} we report the partial structure factors calculated from simulation and the $S(q)$ obtained via the fit of the direct correlation function in reciprocal space. We observe that asymptotic law is fully consistent with the data behaviour for all molarity studied.

Isothermal compressibility and partial molar volume are given in fig.\ref{fig:si-kvmnacl}. Isothermal compressibility determined via KB inversion are compatible to our values found using volume fluctuations in NPT ensemble and those reported in ref.~\citealt{weerasinghe2003jpcb}. The isothermal compressibility is also in very good agreement with the experimental values\cite{chitraMolecularAssociationSolution2002} used to establish the force field. Partial volume found are nevertheless close to the values of ref.~\cite{weerasinghe2003jcp} but differs substantially from the experimental values \cite{chitraMolecularAssociationSolution2002}.

\clearpage

\bibliography{Biblio_KBI}